\documentclass[aps,prd, twocolumn,nofootinbib,groupedaddress,superscriptaddress]{revtex4-2}
\usepackage{newtxtext,newtxmath}
\usepackage[normalem]{ulem}
\usepackage[T1]{fontenc}
\DeclareRobustCommand{\VAN}[3]{#2}
\let\VANthebibliography\thebibliography
\def\thebibliography{\DeclareRobustCommand{\VAN}[3]{##3}\VANthebibliography}

\usepackage{graphicx}	
\usepackage{amsmath}	
\usepackage{esint} 
\usepackage{xcolor}
\usepackage{mathtools}
\usepackage{hyperref} 	
\usepackage[all]{hypcap}
\hypersetup{
    colorlinks=true,	%false: black links; true: colored links
    linkcolor=blue,		%color of internal links
    citecolor=blue,		%color of links to bibliography
    filecolor=magenta,	%color of file links
    urlcolor=blue	    %color of external links
}

\newcommand{\eref}[1]{eq.~\eqref{#1}}

\newcommand{\erefs}[2]{eqs.~\eqref{#1}~and~\eqref{#2}}

\newcommand{\sref}[1]{sec.~\ref{#1}}

\newcommand{\aref}[1]{app.~\ref{#1}}

\newcommand{\fref}[1]{fig.~\ref{#1}}

\newcommand{\Rref}[1]{Ref.~\cite{#1}}
\newcommand{\rref}[1]{ref.~\cite{#1}}

\newcommand{\rrefs}[1]{refs.~\cite{#1}}

\providecommand{\abs}[1]{\lvert#1\rvert}
\providecommand{\bd}[1]{\boldsymbol{#1}}
\providecommand{\ro}[1]{\mathrm{#1}}

\newcommand{\boldB}{\boldsymbol{B}}
\newcommand{\boldE}{\boldsymbol{E}}
\newcommand{\boldF}{\boldsymbol{F}}
\newcommand{\boldv}{\boldsymbol{v}}
\newcommand{\boldj}{\boldsymbol{j}}
\newcommand{\boldr}{\boldsymbol{r}}
\newcommand{\boldp}{\boldsymbol{p}}
\newcommand{\bolde}{\boldsymbol{e}}
\newcommand{\boldnabla}{\boldsymbol{\nabla}}
\newcommand{\gmax}{\bar{\gamma}}

\newcommand{\Je}{J_{e}}
\newcommand{\boldje}{\boldj_{e}}
\newcommand{\Jm}{J_{m}}
\newcommand{\boldjm}{\boldj_{m}}

\newcommand{\JM}{J_{\scriptscriptstyle{\text{M}}}}
\newcommand{\JMb}{J_{\bar{\scriptscriptstyle{\text{M}}}}}

\newcommand{\nM}{n_{\scriptscriptstyle{\text{M}}}}
\newcommand{\nMb}{{n_{\bar{{\scriptscriptstyle{\text{M}}}}}}}

\newcommand{\boldvM}{\boldv_{\scriptscriptstyle{\text{M}}}}
\newcommand{\boldvMb}{\boldv_{\bar{\scriptscriptstyle{\text{M}}}}}
\newcommand{\gammaM}{\gamma_{\scriptscriptstyle{\text{M}}}}
\newcommand{\gammaMb}{\gamma_{\bar{\scriptscriptstyle{\text{M}}}}}

\newcommand{\fM}{f_{\scriptscriptstyle{\text{M}}}}
\newcommand{\fMb}{f_{\bar{\scriptscriptstyle{\text{M}}}}}

\begin{document}

%===============
%  Title 
%===============

\title{Magnetic monopole plasma oscillations and implications for TeV blazars}

%===============

\author{Mariia Khelashvili}
\email{mariia.khelashvili@sissa.it}
\affiliation{SISSA, International School for Advanced Studies, via Bonomea 265, 34136 Trieste, Italy}
\affiliation{INFN Sezione di Trieste, via Valerio 2, 34127 Trieste, Italy}
\author{Takeshi Kobayashi}
\email{takeshi.kobayashi@sissa.it}
\affiliation{SISSA, International School for Advanced Studies, via Bonomea 265, 34136 Trieste, Italy}
\affiliation{INFN Sezione di Trieste, via Valerio 2, 34127 Trieste, Italy}
\affiliation{IFPU, Institute for Fundamental Physics of the Universe, via Beirut 2, 34014 Trieste, Italy}
\affiliation{Kobayashi-Maskawa Institute for the Origin of Particles and the Universe, Nagoya University, Nagoya 464-8602, Japan}
\affiliation{Kavli Institute for the Physics and Mathematics of the Universe (WPI), University of Tokyo, Kashiwa 277-8583, Japan}
\author{Andrew J.~Long}
\email{andrewjlong@rice.edu}
\affiliation{Department of Physics and Astronomy, Rice University, Houston, Texas 77005, U.S.A.}

%\date{Accepted XXX. Received YYY; in original form ZZZ}

\begin{abstract}
    Magnetic monopoles arise in many beyond Standard Model scenarios, symmetrize Maxwell's equations, and their existence would be tied to the quantization of electric charge. It has been argued that, when placed in an astrophysical magnetic field, monopoles can induce a magnetic version of plasma oscillations.
    In this work, we explore monopole-induced oscillations of the intergalactic magnetic field (IGMF). We show that monopole-induced oscillations of the magnetic field lead to ``collimation'' of electrically charged particle trajectories, reducing the usual deflection by the magnetic field. The collimation effect impacts the deflection angle in the electromagnetic cascades of TeV blazars and leads to a decrease in the angular size of blazar secondary GeV halos. Therefore, the constraints on the secondary halo angular size from combined H.E.S.S. and Fermi-LAT observations translate into bounds on the magnetic monopole abundance. The bounds on the magnetic monopole flux obtained in this work from blazar 1ES 0229+200, depending on the IGMF strength, can be as strong as $F \lesssim 6 \times 10^{-23}\, \text{cm}^{-2} \text{s}^{-1} \text{str}^{-1}$ for low-mass monopoles $m \lesssim 10^6\, \text{GeV}$, stronger than existing laboratory and astrophysical bounds. The bound becomes subdominant to current constraints if the present-day IGMF value is stronger than $B \gtrsim 10^{-12}$~G. At the same time, in the case of non-zero monopole abundance, the IGMF lower bound from TeV observations itself should be revised, resulting in a stronger lower bound at higher monopole number density.
\end{abstract}

\date{\today}

\maketitle

\tableofcontents

\newpage

%===============
%  Introduction
%===============

\section{Introduction}

The hypothetical magnetically charged particles known as magnetic monopoles would symmetrize Maxwell’s equations. Moreover their existence would be tied to the electric charge quantization, as manifested in the Dirac quantization condition on the product of electric and magnetic charges, 
$e g = 2\pi N$, with $N$ being an integer~\cite{Dirac:1931kp}.
Moreover, monopoles arise in various beyond-Standard-Model scenarios.
In particular in grand unified theories, they are predicted to exist 
as topological solitons ('t~Hooft--Polyakov monopoles \cite{tHooft:1974kcl, Polyakov:1974ek}). See e.g. \cite{Preskill1984,Milton_2006,Weinberg:2006rq,Rajantie:2012xh,Mavromatos:2020gwk,ParticleDataGroup:2024cfk,Mitsou:2026zcf} for reviews on monopoles from different perspectives.

In astrophysical settings, the presence of monopoles leads to a magnetic analogue of plasma oscillations~\cite{Parker1987}. Depending on the properties of the astrophysical medium and the monopole parameters, 
the oscillations quickly damp and result in the dissipation of magnetic fields. Consequently, the survival of astrophysical magnetic fields imposes bounds on the monopole abundance. 
Monopoles in the galaxy are thus constrained by the galactic Parker bound~\cite{Parker:1970xv,Turner_1982}, derived from the observed present-day galactic magnetic field with strength $B \sim 10^{-6}$~G and correlation length of $\lambda_B \sim 1$~kpc, and the seed Parker bound, which is based on the survival of seed galactic magnetic  fields~\cite{Adams:1993fj}.

In contrast, monopoles in the intergalactic space induce coherent oscillations of the intergalactic magnetic field (IGMF)~\cite{Long:2015bda, Kobayashi_2023,Perri_2024}. 
For most of the relevant monopole parameter space, these oscillations do not decay; instead, the IGMF can continue to oscillate with a nearly constant amplitude over a cosmological timescale. 
Consequently, Parker-like arguments based on magnetic field survival cannot be employed for the present-day IGMFs.
At the same time, magnetic field oscillations can manifest in observations probing the IGMF strength. In this work, we show that monopole-induced oscillations affect the morphology of electromagnetic cascades from TeV blazars, thereby enabling new constraints on the monopole abundance.

Observations of TeV blazars have been interpreted as evidence for an IGMF~\citep{Neronov_2010, Takahashi_2011, HESS_2014, Archambault_2017, Ackermann_2018, Tiede_2020, HESS:2023zwb, Dermer:2010mm}. Although alternative hypotheses for the blazar observations exist~\cite{Broderick_2012, Broderick_2014, Chang_2014, Chang_2016, Shalaby_2018, Broderick_2018, Shalaby_2020, Lamberts_2022, Blanco:2023kfa}, the IGMF explanation is currently widely regarded as a leading interpretation. Such an IGMF could also serve as a seed for the galactic magnetic field, later amplified by dynamo mechanisms during galaxy formation. 
In this work, we assume the existence of an IGMF with strength in the range $B \sim 10^{-15} - 10^{-9}$~G, as inferred from blazar observations and cosmic microwave background (CMB) anisotropies~\cite{Planck:2015zrl,Zucca:2016iur,Jedamzik:2018itu}.
We focus in particular on IGMFs with large correlation lengths, $\lambda_B \gg 100$~kpc.

\begin{figure*}[t]
    \centering
    \includegraphics[width=\linewidth]{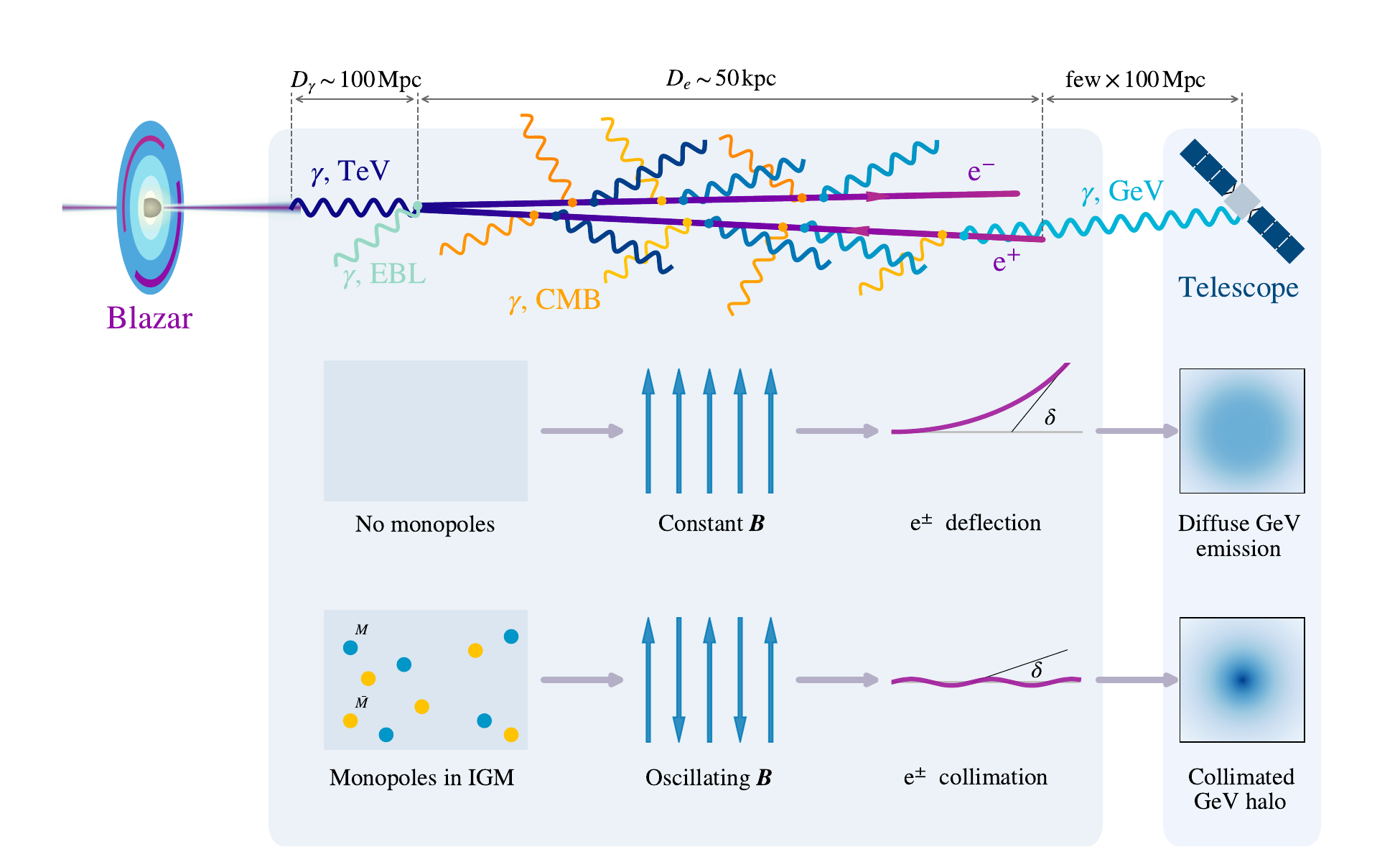}
    \caption{
    Illustration of magnetically-broadened blazar cascade.  An active galactic nucleus emits TeV $\gamma$ rays toward telescopes on Earth.  Interactions of the gamma rays with extragalactic background light (EBL) and cosmic microwave background (CMB) radiation leads to electron/positron pair creation and the development of an electromagnetic cascade (the distances in the plot are not to scale). 
    The presence of an approximately static intergalactic magnetic field (IGMF) deflects the $e^-$ / $e^+$ trajectories and broadens the cascade, leading to a diffuse GeV gamma ray halo.  A sufficiently high density of magnetic monopoles induces oscillations in the magnetic field, leading to reduced deflection (i.e., collimation), and a less diffuse halo.  
    }
    \label{fig:blazar-exeperiment-cartoon}
\end{figure*}

The impact of monopole-induced IGMF oscillations on TeV blazar observations is the main focus of this work. We model the development of electromagnetic (EM) cascades from TeV blazars in the presence of monopole-induced oscillations of the IGMF.
We find that, under certain conditions, these oscillations can counteract the cascade diffusion by ``collimating'' $e^\pm$ trajectories, and reduce the halo size. The non-observation of the secondary photon halo of TeV blazars can therefore be used to constrain the magnetic monopole abundance, since the collimating effect is larger for higher monopole number densities. 
The EM cascades and observations of TeV blazars are illustrated in Fig.~\ref{fig:blazar-exeperiment-cartoon}. The top row below the sketch shows the classical scenario with a constant IGMF, while the bottom row illustrates how monopole-induced oscillations affect the process.  

%--------------------
%  Fig 1
%--------------------

Throughout this article, we adopt the following conventions. 
The Minkowski metric is denoted by $\eta_{\mu\nu}$, and it is normalized as $\eta_{\mu\nu} = (1,-1,-1,-1)$.  
Greek letters are used for the spacetime indices, $\mu, \nu = 0,
1, 2, 3$, and Latin letters for spatial indices, $i, j = 1,2,3$. 
We work in the Heaviside-Lorentz unit system and set $\hbar = c = k_B = 1$.  
The unit of magnetic field is $\mathrm{G} \approx 0.0195 \, \mathrm{eV}^2$. 
The electric charge $e$ is in units of the elementary charge $\bar{e} \equiv \sqrt{4 \pi \alpha} \approx 0.303$, and magnetic charge $g$ is a multiple of the Dirac charge $\bar{g} \equiv 2 \pi / \bar{e} \approx 20.7$. We use for Hubble expansion rate $H_0 = 67.4\, \text{km}/\text{s}/\text{Mpc}$~\cite{Planck2018}. 
 
%===============
%  TeV blazars as a probe of the IGMF
%===============
\section{TeV blazars as a probe of the IGMF}
\label{sec:blazar}

%=====
Observations of TeV blazars furnish a powerful probe of the IGMF. 
Blazars are active galactic nuclei, which  accelerate charged particles to relativistic speeds, resulting in an electromagnetic spectrum that reaches energies as high as $\sim10$ TeV.  
Photons of such energy cannot propagate for cosmological distances through extragalactic space; instead, they scatter on the extragalactic background light (EBL) and produce electron-positron $e^{\pm}$ pairs, which in turn upscatter CMB photons and emit inverse-Compton (IC) radiation, developing an electromagnetic cascade. 
This cascade is expected to generate a secondary $\sim$GeV photon spectrum from the blazar by converting initial TeV photons into GeV radiation. 
Yet, the expected contribution from secondary GeV emission is not observed in the spectrum of TeV blazars. 
The most widely accepted interpretation is that $e^\pm$ pairs are deflected off of the line of sight by the IGMF, which enables one to derive a lower limit on the IGMF strength~\cite{Neronov_2009}. 
An alternative explanation for the missing secondary photons is 
the development of plasma beam instabilities within the cascade~\cite{Broderick_2012}; this hypothesis however has recently been questioned in~\cite{Arrowsmith:2025apl}.
If the GeV gamma ray halo were detected with future observations, it would yield evidence for the IGMF interpretation, and measurements of the halo's size, photon arrival time delay, and morphology would furnish information about the IGMF's strength, coherence length, and helicity~\cite{Long:2015bda,Broderick:2016akd,Tiede:2017xng,Stern:2025shk}.

In the remainder of this section, we briefly summarize the formation of a blazar cascade halo following \rref{Neronov_2010}.  
In the following sections, we discuss the impact of magnetic monopoles on the angular extent of the halo and their implications for limits on the IGMF field strength.

%--------------------
%  Blazar-induced electromagnetic cascade
%--------------------
\subsection{Blazar-induced electromagnetic cascade}

%=====
The blazar's intrinsic spectrum contains hard ``primary'' photons with energy $E_{\gamma,p} \sim \mathrm{TeV}$.  
Such energetic photons are likely to be absorbed by the EBL if they propagate for cosmological distances.  
(That said, in this paper we focus on blazars that are not too distant, i.e. $z \ll 1$, and thus we ignore the expansion of the universe.)
The mean free path of the photons is approximately 
\begin{equation}\label{eq:D_gamma}
    D_{\gamma} \approx 80 \,  \left(\frac{E_{\gamma,p}}{10\, \text{TeV}} \right)^{-1} \text{Mpc}\,, 
\end{equation}
where the authors of \rref{Neronov_2009} discuss an $\mathcal{O}(1)$ uncertainty in this relation that arises from modeling the sources of EBL radiation at cosmological distances. 
Each absorbed TeV gamma ray produces an $e^\pm$ pair with energy 
\begin{equation}
    E_e \approx \frac{E_{\gamma,p}}{2} \,.
\end{equation}
The ultra-relativistic electrons and positrons are likely to scatter elastically on CMB photons within $\sim 100 \, \mathrm{kpc}$.  
This is an inverse-Compton scattering, which produces ``secondary'' gamma rays having an energy of approximately 
\begin{equation}
    E_{\gamma,s} = \frac{4}{3} \epsilon_\text{CMB} \left(\frac{E_e}{m_e}\right)^2 \approx 81\,\left( \frac{E_e}{5\,\text{TeV}}\right)^2 \text{GeV}\,,
    \label{E-gamma-IC} 
\end{equation}
where $\epsilon_{\text{CMB}} = (\pi^4/30\, \zeta(3))\, T_0 \approx 6.3 \cdot10^{-4} \, \text{eV}$ 
is the average energy of a CMB photon,
$T_0$ is the present-day CMB temperature,
and $m_e \approx 0.511 \, \text{MeV}$ is the electron mass.  
Over multiple scatterings, the average energy loss rate is
\begin{equation}
    P_e = \frac{4}{3} \sigma_{T} u_{\text{CMB}} \left(\frac{E_e}{m_e} \right)^2 \,,
\end{equation}
where $\sigma_T$ is the Thomson cross-section constant and $u_{\text{CMB}} = (\pi^2/15) T_0^4 \approx 0.26\, \text{eV} \cdot \text{cm}^{-3}$ 
is the energy density of the CMB.
The characteristic distance on which the $e^\pm$ lose an order one fraction of their energy is thus estimated as 
\begin{equation}
    D_e = \frac{E_e}{P_e} 
    \approx 73\, \left( \frac{E_e}{5\, \text{TeV}} \right)^{-1} \text{kpc} \,.
    \label{De}
\end{equation}

%--------------------
%  Magnetic broadening of the cascade
%--------------------
\subsection{Magnetic broadening of the cascade}

%=====
If the electromagnetic cascade develops in the presence of an intergalactic magnetic field, then the trajectories of the electrons and positrons are bent by the Lorentz force.  
The characteristic length scale of this bending is given by the Larmor radius $R_L$.  
For an ultra-relativistic 
electron or positron moving through an approximately uniform magnetic field, the Larmor radius is given by $R_L = v_{e,\perp} / \omega_c$  where $v_{e,\perp}$ is the component of the particle's velocity normal to the magnetic field and $\omega_c = \bar{e} \bar{B} / E_e$  is the cyclotron frequency, with $\bar{B} = |\boldB|$ the magnetic field strength.  
The magnetic field can be treated as homogeneous if its coherence length $\lambda_B$ is large compared to the distance that the electrons and positrons travel, i.e. $\lambda_B \gg D_e \sim 100 \, \ro{kpc}$.  
As a result of this bending, the GeV gamma rays are deflected off of the line of sight by an angle that can be as large as $\delta \approx D_e / R_L$ for $D_e \ll R_L$.  
Conversely, some gamma rays that would have otherwise not reached Earth are bent back toward the line of sight, 
and consequently the blazar develops a halo of GeV gamma rays. For $D_e > R_L$ the GeV gamma rays are emitted approximately isotropically.

The angular extent~$\theta_{\ro{ext}}$ {(i.e., the angular size as seen by an observer at a distance $D$)} of the halo formed by secondary photons emitted in the electromagnetic cascade is related to the deflection angle~$\delta$ via 
\begin{equation}
 \sin \theta_{\ro{ext}} = 
\frac{ \sin \delta }{\tau }.
\label{eq:sindelta}
\end{equation}
Here $\tau = D / D_{\gamma}$ ($\gtrsim 1$) is the optical depth for gamma rays due to absorption by the EBL, 
$D_\gamma$ is given by \eref{eq:D_gamma},
and we also used that $D_e \ll D_{\gamma}$.
In the next subsection we discuss how blazar observations lead to constraints on the IGMF, and in later sections we demonstrate how the presence of magnetic monopoles modify these constraints via the calculation of $\delta$. An illustration of the TeV blazar EM cascade is shown in \fref{fig:blazar-exeperiment-cartoon}.

%--------------------
%  Blazar observations and IGMF constraints
%--------------------
\subsection{Blazar observations and IGMF constraints}
\label{sec:blazar-fiducail}

%=====
Multi-band observations of several TeV blazars do not detect the GeV gamma ray flux that is expected to arise through the electromagnetic cascade.  
These observations can be explained by the presence of an intergalactic magnetic field, which broadens the cascade and reduces the flux of GeV gamma rays reaching Earth.  
A lower limit on the magnetic field strength can be derived by comparing the angular extent of the halo $\theta_\text{ext}$ with the gamma ray telescope's point spread function (PSF) $\theta_\text{psf}$, imposing
\begin{align}\label{eq:ext_vs_psf}
     \theta_\text{ext} \geq \theta_\text{psf} 
     \;.
\end{align}
The first such limits were derived by the authors of \rref{Neronov_2010} (and simultaneously \cite{Tavecchio:2010mk}) using a combination of H.E.S.S. and Fermi-LAT spectra.  
Their strongest bound was obtained for the blazar 1ES 0229+200, which is cosmologically distant ($z \approx 0.14$) with the line of sight to it lying mostly within voids and possesses a hard spectrum extending up to $15 \, \mathrm{TeV}$ \cite{Stecker:2007jq,Aliu:2013pya}. 
The expected flux from the EM cascade exceeds the Fermi-LAT limits below $E_{\gamma, \text{min}} \approx 39$~GeV (assuming a power-law index for the intrinsic blazar spectrum of $\Gamma \ge 1.5$).  
The PSF of Fermi-LAT at this energy scale is $\theta_\text{psf} \sim 0.2^\circ$, and the optical depth for $E_{\gamma,p} \sim 1$--$10$~TeV photons from a source at redshift $z\sim 0.1$ is $\tau \sim 1$--$4$~\cite{Dominguez_2010}.
Since the product of the PSF and optical depth satisfies
$\tau \theta_{\ro{psf}} \ll 1 $,
the condition (\ref{eq:ext_vs_psf}),
together with (\ref{eq:sindelta}) expanded to linear order in the angles, 
yield a lower limit on the deflection angle,
\begin{equation}
 \delta \geq \tau \theta_{\ro{psf}}.
\label{eq:deltaLB}
\end{equation}
Then using $\delta \approx D_e / R_L \propto \bar{B}$, 
a lower bound on the IGMF was obtained as $B \geq 3\times 10^{-16}$~G for $\lambda_B \gtrsim 0.1$~Mpc \cite{Neronov_2010}. 

Subsequent studies~\cite{Takahashi_2011, HESS_2014, Archambault_2017, Ackermann_2018, Tiede_2020, HESS:2023zwb} refined and extended this analysis, yielding more stringent and systematic constraints.  
A recent study \cite{HESS:2023zwb} reports a conservative bound on the IGMF strength at the level of $B \ge 7.1 \times 10^{-16} \, \mathrm{G}$ for $\lambda_B = 1 \, \mathrm{Mpc}$, and under less conservative assumptions about the blazar duty cycle, the limits strengthen by a factor of $25$ to $50$.  A complementary limit of $B \gtrsim 10^{-18}$~G was derived in \cite{Dermer:2010mm} through an analysis of cascade radiation time delays and is independent of variations in the intrinsic blazar spectrum.

%===============
%  Magnetic Langmuir oscillations
%===============
\section{Magnetic Langmuir oscillations}
\label{sec:monopoles-electrodynamics}

%=====
In this section we discuss how a system of magnetic charges in a homogeneous magnetic field will exhibit oscillatory dynamics~\cite{Parker1987,Long:2015cza,Perri_2024}, which is a magnetic analog of Langmuir oscillations.  
We then study the effect of these magnetic Langmuir oscillations on the trajectory of an electric charge, and we assess the impact on the magnetic broadening of a TeV blazar's electromagnetic cascade.  

%--------------------
%  Electrodynamics with magnetic charges
%--------------------
\subsection{Electrodynamics with magnetic charges}

%=====
We consider an extension of classical electromagnetism that includes magnetic monopoles.  
In presence of monopoles, Maxwell's equations are modified to 
\begin{equation}\label{eq:covMax}
    \begin{cases}
        \partial_{\mu} F^{\mu\nu} = \Je^{\nu} \\
        \partial_{\mu} \tilde{F}^{\mu\nu} = \Jm^{\nu}
    \end{cases}
\end{equation}
where $F_{\mu\nu}$ is the electromagnetic field strength tensor field, and $\tilde{F}^{\mu\nu} = \frac{1}{2} \epsilon^{\mu\nu\rho\sigma} F_{\rho\sigma}$ is its dual,
with $\epsilon^{\mu\nu\rho\sigma}$ being the totally-antisymmetric Levi-Civita symbol normalized as $\epsilon^{0123} = +1$.  
Moreover, $\Je^\mu(x)$ is the familiar electric current density 4-vector field, and $\Jm^\mu(x)$ is the new magnetic current density 4-vector field.  
In 3-vector notation, these equations are expressed as 
\begin{equation}
    \begin{cases}
        \boldnabla \cdot \boldE = \rho_e \\
        \boldnabla \cdot \boldB = \rho_m \\
        \boldnabla \times \boldE = -\frac{\partial \boldB}{\partial t} - \boldjm \\
        \boldnabla \times \boldB = \frac{\partial \boldE}{\partial t} + \boldje
    \end{cases} \;.
    \label{maxwell-equations}
\end{equation}
Here the electromagnetic fields are given by
$E^i = F^{i0}$ and 
$B^i = -\frac{1}{2}\varepsilon^{ijk} F_{jk} = \tilde{F}^{i0}$
with 
$\varepsilon^{ijk} = \epsilon^{0 i jk}$,
and the electromagnetic charge and current densities are 
$\rho = J^0$ and $j^i = J^i$.   
It is clear from the covariant equations (\ref{eq:covMax}) that
both the electric and magnetic currents are required to obey continuity equations of the form $\partial_\mu J^\mu = \tfrac{\partial \rho}{\partial t} + \boldnabla \cdot \boldj = 0$, which express the local conservation of electric and magnetic charge.  

%=====

An electrically charged particle with charge $e$ and mass $m_e$,
and a magnetically charged particle with charge $g$ and mass $m_m$, 
respectively move in response to an applied electromagnetic field according to the equations of motion
\begin{subequations}
\begin{align}
m_e \tfrac{\mathrm{d}}{\mathrm{d}\tau} U^{\mu} & = e F^{\mu \nu} U_\nu 
    \;,
\\
m_m \tfrac{\mathrm{d}}{\mathrm{d}\tau} U^{\mu} & = g \tilde{F}^{\mu \nu} U_\nu 
    \;.
    \label{eom-covariant}
\end{align}
\end{subequations}
Here $X^\mu(\tau)$ is the particle's world line, $U^\mu(\tau) = \tfrac{\mathrm{d}}{\mathrm{d}\tau} X^\mu$ is the particle's 4-velocity,
and $\tau$ is the particle's proper time with normalization $\eta_{\mu\nu} U^\mu U^\nu = 1$ for all $\tau$.  
In a frame where the particle's 3-velocity is $\boldv$, the velocity may be written as $U^\mu(\tau) = (\gamma, \gamma \boldv)$
where $\gamma = 1 / \sqrt{1 - |\boldv|^2}$ is the Lorentz factor.  
Along the particle's worldline, the time coordinate is $t = X^0(\tau)$, which implies $\mathrm{d}/\mathrm{d}\tau = \gamma(t) \mathrm{d}/\mathrm{d}t$.  
In 3-vector notation, the time and spatial components of the 
equations of motion are expressed as 
\begin{subequations}
\begin{align}
&\begin{cases}
        m_e \tfrac{\mathrm{d}}{\mathrm{d}t} \gamma  = e \boldE \cdot \boldv  \,, \\ 
        m_e \tfrac{\mathrm{d}}{\mathrm{d}t} (\gamma \boldv)  = e \boldE + e \boldv \times \boldB \,,
\end{cases}
\label{eq:elecEOM}
\\
&\begin{cases}
        m_m \tfrac{\mathrm{d}}{\mathrm{d}t} \gamma = g \boldB \cdot \boldv \,, \\ 
        m_m \tfrac{\mathrm{d}}{\mathrm{d}t} (\gamma \boldv) = g \boldB - g \boldv \times \boldE \,.
\end{cases}
\label{eom-noncovariant}
\end{align}
\end{subequations}
Note that in both sets of equations, 
the first line follows from the second line since 
$\ro{d}\gamma / \ro{d}t = \bd{v} \cdot \ro{d}(\gamma \bd{v}) / \ro{d}t$.

%--------------------
%  System of monopoles and anti-monopoles
%--------------------
\subsection{System of monopoles and anti-monopoles}
\label{sec:monopoles}

%=====
Suppose that there exist 
monopoles $\ro{M}$ with magnetic charge $g$
and anti-monopoles $\bar{\ro{M}}$ with charge $-g$, 
with equal mass $m$.
Hereafter we take $g > 0$ without loss of generality. 
We are interested in a system containing many such monopoles and anti-monopoles interacting with the electromagnetic field.  
By performing a spatial coarse-graining on a scale that is much larger than the typical separation between monopoles and anti-monopoles, we are able to characterize the system with continuous density and (bulk) velocity fields.  
(We are implicitly assuming the monopoles and anti-monopoles to each move somewhat coherently.)
The monopoles and anti-monopoles each have a 4-vector current density, denoted by $\JM^\mu(x)$ and $\JMb^\mu(x)$.  
We assume that monopoles and anti-monopoles are neither created nor destroyed, 
which implies that the currents are separately conserved,
\begin{equation}
\partial_\mu \JM^\mu = \partial_\mu \JMb^\mu = 0. 
\label{eq:current-cons}
\end{equation}
The total magnetic current density is 
$\Jm^\mu = \JM^\mu + \JMb^\mu$. 
  
%=====

Moving to 3-vector notation, we represent the monopole number densities as $\nM(\boldr,t)$ and $\nMb(\boldr,t)$, and the velocity fields as $\boldvM(\boldr,t)$ and $\boldvMb(\boldr,t)$, which imply\footnote{The 4-current is related to the 4-velocity field via
$\JM^\mu = g 
n_{\scriptscriptstyle{\text{M}0}}
U_{\scriptscriptstyle{\text{M}}}^\mu
$, where $n_{\scriptscriptstyle{\text{M}0}}$ is the number density of the monopoles in their rest frame.} 
$\JM^\mu = g \nM (1, \boldvM)$ and
$\JMb^\mu = -g\nMb(1, \boldvMb)$.
The total magnetic charge density and magnetic current density are calculated as $\rho_m = g (\nM - \nMb)$ and $\boldjm = g (\nM \boldvM - \nMb \boldvMb)$.   
In the absence of any other charged matter, Maxwell's equations  \eqref{maxwell-equations} reduce to 
\begin{equation}\label{eq:Max13}
    \begin{cases}
        \boldnabla \cdot \boldE = 0 \,, \\
        \boldnabla \cdot \boldB = g (\nM - \nMb) \,, \\
        \boldnabla \times \boldE = -\frac{\partial \boldB}{\partial t} -  g (\nM \boldvM - \nMb \boldvMb) \,, \\
        \boldnabla \times \boldB = \frac{\partial \boldE}{\partial t} \,. 
    \end{cases}
\end{equation}
Moreover, the conservation of monopole and anti-monopole numbers~(\ref{eq:current-cons}) are rewritten as
\begin{equation}\label{eq:cons14}
\begin{split}
\tfrac{\partial}{\partial t} \nM + \boldnabla \cdot (\nM \boldvM )
 & = 0 \,, \\ 
\tfrac{\partial}{\partial t} \nMb + \boldnabla \cdot (\nMb \boldvMb )
 & = 0 \,.
\end{split}
\end{equation}
The evolution equations of the velocity fields take the same form
as the equations of motion of the individual monopoles.
Rewriting the Lagrangian derivative as 
$\ro{d} / \ro{d} t = \partial / \partial t + \bd{v} \cdot \boldnabla$,
the second line of \eqref{eom-noncovariant} yields, 
\begin{equation}\label{eq:vel16}
\begin{split}
    m \left( \tfrac{\partial}{\partial t} + \boldvM \cdot \boldnabla \right)
 \gammaM \boldvM & = g \boldB - g \boldvM \times \boldE \,, \\ 
    m \left( \tfrac{\partial}{\partial t} + \boldvMb \cdot \boldnabla
\right)
 \gammaMb \boldvMb & = - g \boldB + g \boldvMb \times \boldE \,.
\end{split} 
\end{equation}
We assume that other forces (such as gravity) are negligible compared to the electromagnetic force. 
Also, given the rather strict limits on the monopole abundance in astrophysical environments, we ignore the interactions between the individual monopoles and treat them as effectively collisionless. This treatment is justified in App.~\ref{app:C2}.   

%=====
The set of equations (\ref{eq:Max13}), (\ref{eq:cons14}), and (\ref{eq:vel16}) admit solutions with vanishing electric field $\boldE(\boldr,t) = 0$; homogeneous magnetic field $\boldB(\boldr,t) = \boldB(t)$; equal, homogeneous, and static densities $\nM(\boldr,t) = \nMb(\boldr,t) \equiv n/2$ (thus $n$ denotes the total number density of monopoles and anti-monopoles); and anti-aligned and homogeneous velocities $\boldvM(\boldr,t) = - \boldvMb(\boldr,t) \equiv \boldv(t)$ and $\gammaM(\boldr,t) = \gammaMb(\boldr,t) \equiv \gamma(t)$.  
These homogeneous fields trivially satisfy most of the evolution equations, except for the third line of (\ref{eq:Max13}) 
and (\ref{eq:vel16}), which reduce to
\begin{align}
    \frac{\mathrm{d} \boldB}{\mathrm{d} t} 
    & = - g n \boldv \,, 
    \label{maxwell} \\ 
    m \frac{\mathrm{d}}{\mathrm{d}t} \left(\gamma \boldv \right) 
    & = g \boldB \,.
    \label{mm-eom}
\end{align}
These equations allow for oscillatory solutions.

%--------------------
%  Fig 2
%--------------------
\begin{figure*}[t!]
    \centering
    \includegraphics[width=\linewidth]{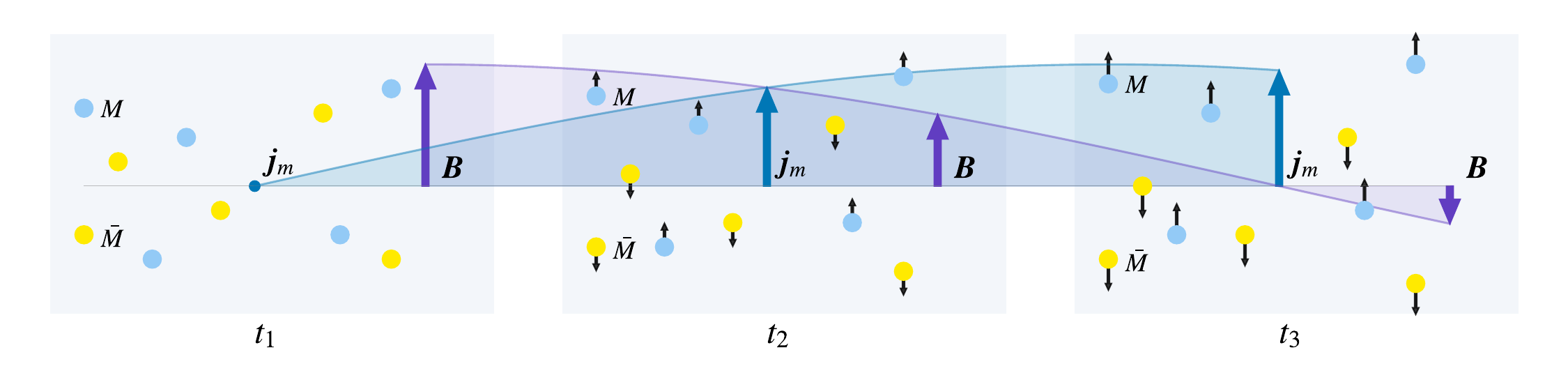}
    \caption{\label{fig:cartoon}
    Illustration of magnetic Langmuir oscillations.  At the first time ($t = t_1$) a population of monopoles ($\scriptstyle{\text{M}}$) and anti-monopoles ($\bar{\scriptstyle{\text{M}}}$) are at rest in a uniform magnetic field ($\boldB$). At the second time ($t = t_2$), the magnetic field accelerates the monopoles and anti-monopoles in opposite directions, leading to a magnetic current density ($\boldjm \propto \boldB$).  At the third time ($t = t_3$), the current sources a magnetic field that opposes the initial magnetic field, causing the total magnetic field to reverse orientation.  Subsequently, the magnetic field decelerates the monopoles, bringing them back to rest, and the system continues to oscillate.  
    }
\end{figure*}

%--------------------
%  Magnetic field oscillations
%--------------------
\subsection{Magnetic field oscillations}
\label{sec:magn-oscillations}

%=====
Taken together, equations~\eqref{maxwell}~and~\eqref{mm-eom} govern the co-evolution of the magnetic field $\boldB(t)$ and the monopole velocity $\boldv(t)$. 
This behavior is illustrated in \fref{fig:cartoon}.  
The physical system is comparable to the phenomenon of plasma oscillations (Langmuir waves) in a conducting medium, and the mathematics is familiar.  
However, we take care to retain the Lorentz factor and thereby account for the possibility that the monopoles and anti-monopoles reach relativistic velocities.  

%=====
If the cosmological expansion can be neglected (i.e., on time scales that are short compared to the age of the universe), then energy is conserved as the system evolves.  
Energy is traded off between the kinetic energy of the monopoles' bulk motion and the magnetic energy, while the sum remains constant: 
\begin{equation}
    m n (\gamma - 1) + \tfrac{1}{2} |\boldB|^2 = \text{const.} \equiv m n (\bar{\gamma} - 1) \equiv \tfrac{1}{2} \bar{B}^2 \,. 
    \label{energy-conserv}
\end{equation}
Here, $\bar{\gamma} = 1/\sqrt{1-\bar{v}^2}$ is the maximal Lorentz factor reached by monopoles during the oscillation period,
while $\bar{B}$ and $\bar{v}$ respectively are the oscillation amplitudes of the magnetic field and monopole velocity.

%=====
We assume that the velocity component of monopoles/anti-monopoles 
perpendicular to $\bd{B}$ is initially zero,
so that in the absence of non-magnetic forces the velocity remains parallel to the magnetic field, i.e.,  $\boldv \parallel \boldB$.
We thus take the oscillations to be in the $z$~direction,
and denote the oscillating components of the magnetic field and velocity by
\begin{equation}
\boldB(t) = B(t) \bolde_z,
\quad
\bd{v}(t) = v(t) \bolde_z. 
\end{equation}
Then, combining equations~\eqref{maxwell}~and~\eqref{mm-eom} yields  
an oscillator equation for the magnetic field
with a time-dependent frequency \citep{Turner_1982,Parker1987,Long:2015cza,Perri_2024}
\begin{equation}\label{eq:ddB}
 \frac{\mathrm{d}^2 B}{\mathrm{d} t^2} + \frac{\Omega_0^2}{\gamma (t)^3}
 B = 0 \,, 
\end{equation} 
where
\begin{equation}
 \Omega_0 = g \sqrt{\frac{n }{m} }.
\end{equation}

%=====
If the parameters are chosen such that the monopoles maintain nonrelativistic speeds (i.e., $\gamma(t) \approx 1$, or equivalently $\bar{B}^2 \ll mn$), then the solution is approximately harmonic 
\begin{subequations}\label{b-harmonic}
\begin{align}
    B(t) & = \bar{B} \cos \Omega_0 t \;, \\  
    v(t) & = \bar{v} \sin \Omega_0 t \;,
\end{align}
\end{subequations}
where $\bar{v} = \bar{B} / \sqrt{m n}$.   
Note that we chose the origin of time so that $B = \bar{B}$ at $t = 0$, and $T = 2 \pi / \Omega_0 $ gives the period of the oscillations.  

Alternatively, if the 
monopoles reach relativistic speeds (i.e., $\gmax \gg 1$ and $\bar{B}^2 \gg mn$), then $v$ switches between $\bar{v} \approx 1$ and $-\bar{v} \approx -1$ every half period of the oscillations.
During a half period while $v$ is constant, 
(\ref{maxwell}) can be solved to give a linear time dependence to $B$, which combined with (\ref{mm-eom}) yields $\gamma$ as a quadratic function of time. 
By patching together the approximate solutions for each half period, one obtains
\begin{subequations}\label{b-triangular}
\begin{align}
    B(t) & = 
    \begin{cases}
        \bar{B} \left(1 - 4\frac{t}{T} \right)\,, &  t \in [0, \frac{T}{2}] \\
        \bar{B} \left(4\frac{t}{T} - 3 \right)\,, &  t \in [\frac{T}{2}, T]
    \end{cases} \;, \\ 
    v(t) & = 
    \begin{cases}
         \bar{v}  \,, &  t \in [0, \frac{T}{2}] \\
        - \bar{v} \,, &  t \in [\frac{T}{2}, T]
    \end{cases} \;, \\
    \gamma(t) &=
    \begin{cases}
 -(\bar{\gamma} - 1) \left(1 - 4\frac{t}{T} \right)^2 + \bar{\gamma}
 \,, &  t \in [0, \frac{T}{2}] \\
 -(\bar{\gamma} - 1) \left(4\frac{t}{T} - 3 \right)^2 + \bar{\gamma}
 \,, &  t \in [\frac{T}{2}, T]
    \end{cases}\;,
\end{align}
\end{subequations}
where 
$\bar{\gamma} = \bar{B}^2 / 2mn + 1$.
The period is $T = 4 \sqrt{2 (\bar{\gamma } - 1)} / \Omega_0$, 
which is much larger than the nonrelativistic value $2 \pi / \Omega_0$
by a factor $ \sim \sqrt{\bar{\gamma}}$.  
 
\begin{figure}
    \centering
    \includegraphics[width=\columnwidth]{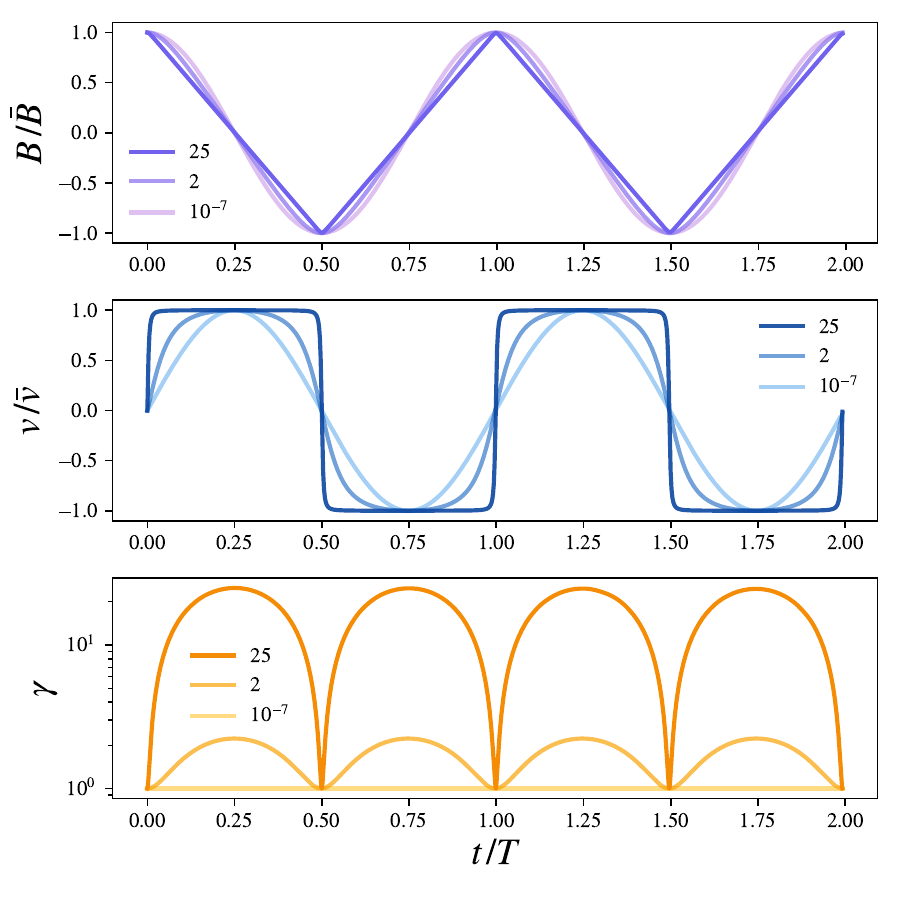}
    \caption{\label{fig:magnetic_field_oscillations}
    Oscillatory solutions for a monopole-magnetic field system.  The magnetic field and monopole velocity evolve according to \erefs{maxwell}{mm-eom}.  We plot solutions of these equations, showing the oscillating components of the magnetic field $B(t)$ and the monopole velocity $v(t)$ normalized by their oscillation amplitudes, as well as the associated Lorentz factor $\gamma(t)$.  
    The scaled solution depends only on a single combination of parameters and for these examples we use $\overline{\gamma v} = 25$, $2$, and $10^{-7}$ as indicated in the plot legends, representing ultra-relativistic, mildly relativistic and non-relativistic oscillations.  
    In the ultra-relativistic regime, monopoles (and counter-propagating anti-monopoles) rapidly reach relativistic speeds, and the magnetic field develops a triangular time profile. In the non-relativistic regime, the oscillations of both the magnetic field and the monopole velocities are harmonic.
    } 
\end{figure}

%=====
The various oscillatory behaviors, including a mildly relativistic case, 
are illustrated in \fref{fig:magnetic_field_oscillations}.  
We show numerical solutions of equations~\eqref{maxwell}~and~\eqref{mm-eom} for a fiducial parameter set in which the 
Lorentz factor–velocity product is taken as
$\overline{\gamma v} = 25$, $2$, and $10^{-7}$. 
In general the system oscillates at an angular frequency 
\begin{equation}
    \Omega \approx \frac{\Omega_0}{ \bar{\gamma}^{1/2}} \,, 
    \label{magnetic-plasma-frequency}
\end{equation}
which is defined such that the corresponding oscillation period is $T = 2 \pi / \Omega$.  
This approximate expression is accurate to within a fractional error of $10\%$ for all values of $\bar{\gamma}$;
we show this explicitly in \aref{app:period}.

%--------------------
%  Fig 3
%--------------------

%--------------------
%  Electrically charged particle trajectory
%--------------------
\subsection{Electrically charged particle trajectory}
\label{sec:particle-trajectory}

%=====

By way of understanding how electrons and positrons propagate through the IGMF, we now consider the trajectory of an electric charge in the homogeneously oscillating magnetic field.  
An electron ($e = -\bar{e}$) or positron ($e = \bar{e}$)
with velocity $\boldv_e(t)$ and Lorentz factor $\gamma_e(t) = 1 / \sqrt{1 - |\boldv_e(t)|^2}$ evolves subject to the equations (cf. (\ref{eq:elecEOM}))
\begin{equation}\label{eq:charge_EOM}
    \frac{\mathrm{d}}{\mathrm{d}t} \gamma_e = 0 ,
    \quad 
    m_e \frac{\mathrm{d}}{\mathrm{d} t} ( \gamma_e \boldv_e ) = \mp \bar{e} \boldv_e \times \boldB 
    \;.
\end{equation}
Notice that $\gamma_e$ is time-independent, since a magnetic field does no work on an electrically charged particle.  
The magnetic field $\boldB(t)$ oscillates with amplitude $\bar{B}$ and period $T$ according to \eqref{eq:ddB}.  
It is useful to identify the Larmor frequency
\begin{equation}\label{eq:wc_def}
    \omega_c = 
    \frac{\bar{e} \bar{B}}{m_e \gamma_e } 
    \;,
\end{equation}
which is the characteristic frequency of the cyclotron motion.  
It is also useful to identify the Larmor radius
\begin{equation}
    R_L = \frac{v_{e,\perp}}{\omega_c} 
    \,,
    \label{R-larmor}
\end{equation}
which is the characteristic length scale of the cyclotron motion.  
The trajectory of the charged particle displays two qualitatively different behaviors depending on whether the time scale for the cyclotron orbits, $2 \pi / \omega_c$, is smaller or larger than the time scale for the magnetic field oscillations $T$.  

%--------------------
%  Fig 4
%--------------------
\begin{figure*}[t]
    \centering
    \includegraphics[width = \linewidth]{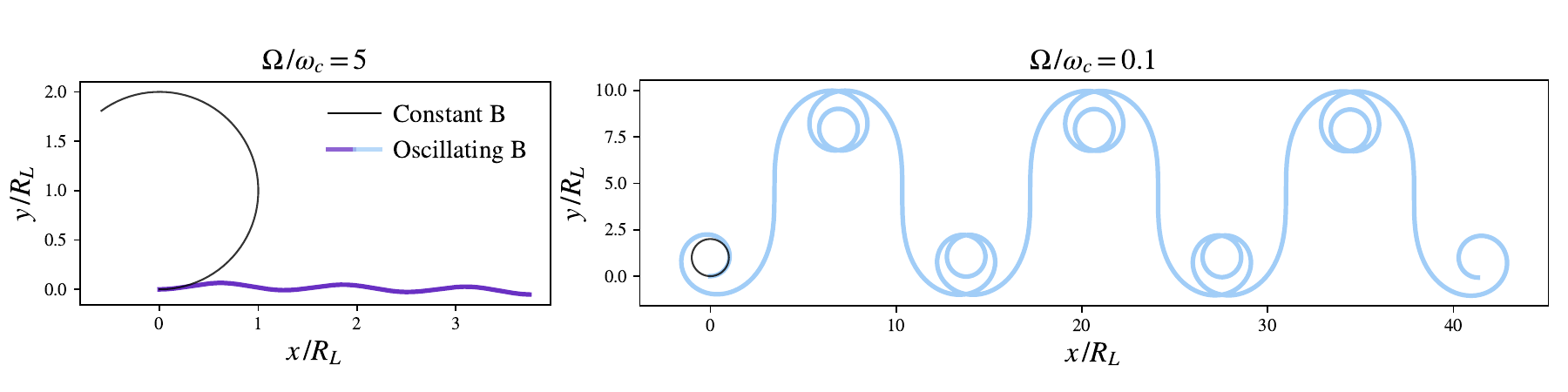}
    \caption{\label{fig:CR_trajectory}
Trajectory of an electrically charged particle in a time-oscillating magnetic field. 
The particle moves according to \eref{eq:charge_EOM}.  
We plot its solution in the $x$-$y$ plane, assuming the magnetic field undergoes ultra-relativistic triangular oscillations along the $z$~direction, and that the particle has a vanishing velocity component parallel to the magnetic field.
Lengths are measured in units of the Larmor radius $R_L$.
    In the two panels we show different choices of the ratio $\Omega / \omega_c$ where $\Omega $ is the magnetic field oscillation frequency and $\omega_c $ is the cyclotron frequency. 
    \textit{\bfseries{Left panel}}: Rapidly oscillating magnetic field with $\Omega / \omega_c = 5$.
    \textit{\bfseries{Right panel}}: Slowly oscillating magnetic field with $\Omega / \omega_c = 0.1$. 
The black line denotes the trajectory in a constant magnetic field.
    }
\end{figure*}

%=====
For systems with $T \ll 2 \pi / \omega_c$, the magnetic oscillation time scale is smaller than the cyclotron time scale, and we can say that the magnetic field is rapidly oscillating.  
An example of the resultant behavior is illustrated in the left panel of \fref{fig:CR_trajectory}.  
The particle's trajectory follows an approximately linear path in its initial direction of motion with small oscillations superimposed.  
Since the magnetic field is rapidly oscillating, the particle feels a much weaker effective field strength.  

%=====
For systems with $2 \pi / \omega_c \ll T$, the cyclotron time scale is smaller than the magnetic oscillation time scale, and we can say that the magnetic field is slowly oscillating.  
An example of the resultant behavior is illustrated in the right panel of \fref{fig:CR_trajectory}.  
When the magnetic field is comparable to its maximum value $|B(t)| \approx \bar{B}$, the particle's trajectory resembles a circular path, which approximates a standard cyclotron orbit.  
However, when the field amplitude is small $|B(t)| \ll \bar{B}$, the particle's trajectory resembles a straight line, which approximates the path of a free particle.  

%--------------------
%  Beaming and reduced deflection
%--------------------

\subsection{Beaming and reduced deflection}
\label{sub:beaming}

Here we estimate the angular deflection that an electron or positron experiences in the presence of the monopole-induced magnetic field oscillations.  

For concreteness let us consider an electron whose velocity component parallel to the magnetic field is initially zero.
Then the electron's trajectory remains in a plane perpendicular to the direction of the magnetic field, and so we write the 
magnetic field and the electron velocity as\footnote{Even in the presence of a parallel velocity component, as long as its amplitude is comparable to or less than that of the perpendicular component, the results for the deflection angle hold at the order-of-magnitude level.}
\begin{equation}
\boldB(t) = B(t) \bolde_z,
\, \, \, \, 
\bd{v}_e(t) = v_{e,\perp} [ \cos \delta (t) \bolde_x + \sin \delta (t) \bolde_y ]. 
\end{equation}
Here note that the velocity's amplitude
$\abs{\bd{v}_e} = v_{e,\perp}$ stays constant, while its direction rotates in the $x$-$y$ plane, which is parametrized by the deflection angle~$\delta(t)$. 
The velocity evolves according to \eref{eq:charge_EOM}, which can be rewritten as
\begin{equation}
 \frac{\mathrm{d} \delta(t)}{\mathrm{d}t} = \frac{\bar{e} B(t)}{m_e \gamma_e}. 
\label{eq:29}
\end{equation}
We choose the origin of time so that $B = \bar{B}$ at $t = 0$, and 
we also set the initial angle to $\delta(0) = 0$. 
Moreover, we parametrize the regularly-oscillating magnetic field as 
\begin{equation}
    B(t) = \bar{B} \, f_B(\tfrac{t}{T}) 
    \;,
\end{equation}
where $f_B(\tau)$ is a periodic function having $f_B(0) = f_B(1) = 1$.  
Then, integrating the equation of motion gives
\begin{equation}
  \delta (t) = \frac{\bar{e}}{m_e \gamma_e} \int^{t}_{0} \mathrm{d}t' \, B(t')
 =  \frac{\omega_c }{ \Omega} \int^{t / T}_{0} \mathrm{d}\tau \, 
2 \pi  f_B(\tau).
\label{eq:28}
\end{equation}
The integral in the far right-hand side 
is maximized at the quarter cycle $t = T/4$, when its value is $1$ for harmonic oscillations \eqref{b-harmonic}, and $\pi/4 \approx 0.785$ for triangular oscillations \eqref{b-triangular}.  
Neglecting this order-unity factor, the deflection angle for $ t \gtrsim T/4$ is thus typically of $\abs{\delta (t)} \sim \omega_c / \Omega$. 
On the other hand for $t \ll T$, the integral gives $\simeq \Omega t$ for both types of oscillations, hence $\delta (t) \simeq \omega_c t$. 

Considering that electrons and positrons in the cascade travel a distance $D_e$ with ultrarelativistic speeds, their total deflection angle can be estimated by substituting $v_{e,\perp} \sim 1$, $t \sim D_e$, and 
$\omega_c \sim 1/R_L$ in the above discussion. 
Hence if $D_e \Omega \gtrsim 1$ we have $\delta \sim 1 / \Omega R_L$.
(Hereafter we remove the modulus sign and use $\delta$ to denote the typical absolute deflection angle.)
On the other hand if $D_e \Omega \lesssim 1$, then we get 
 $\delta \sim D_e / R_L$, which is the same result as in the absence of monopoles. Note that in either case, $\delta$ larger than unity indicates that the electrons/positrons are deflected isotropically; we write the deflection angle in such isotropized cases as $ \delta \sim 1$ (with a slight abuse of notation). 
Then, we can express the deflection angle collectively for all cases as
%The deflection for the two cases can collectively be written as
\begin{equation}
    \delta  \sim \frac{1}{R_L} \min \left(D_e, \, \frac{1}{\Omega} , \, R_L \right) \,.
    \label{deflection-angle}
\end{equation}
We hence clearly see that the deflection is determined by the smallest of the three characteristic length scales
$(D_e, \, \Omega^{-1}, \, R_L)$.
In particular if $R_L < D_e, \Omega^{-1}$,
then the deflection is isotropic.
On the other hand, for the monopole-induced oscillations to suppress the deflection angle, the oscillation scale needs to satisfy 
$\Omega^{-1} <  D_e, R_L$.\footnote{{Note that although the effect is superficially similar to the presence of multiple magnetic-field domains, the magnetic-field oscillations here are coherent. As a result, the deflection angle does not acquire the $\propto \sqrt{N}$ factor that would arise from $N$ domains with randomly oriented magnetic fields, as in a random walk.}}

%--------------------
%  Fig 5
%--------------------
\begin{figure}[t]
    \centering
    \includegraphics[width=\linewidth]{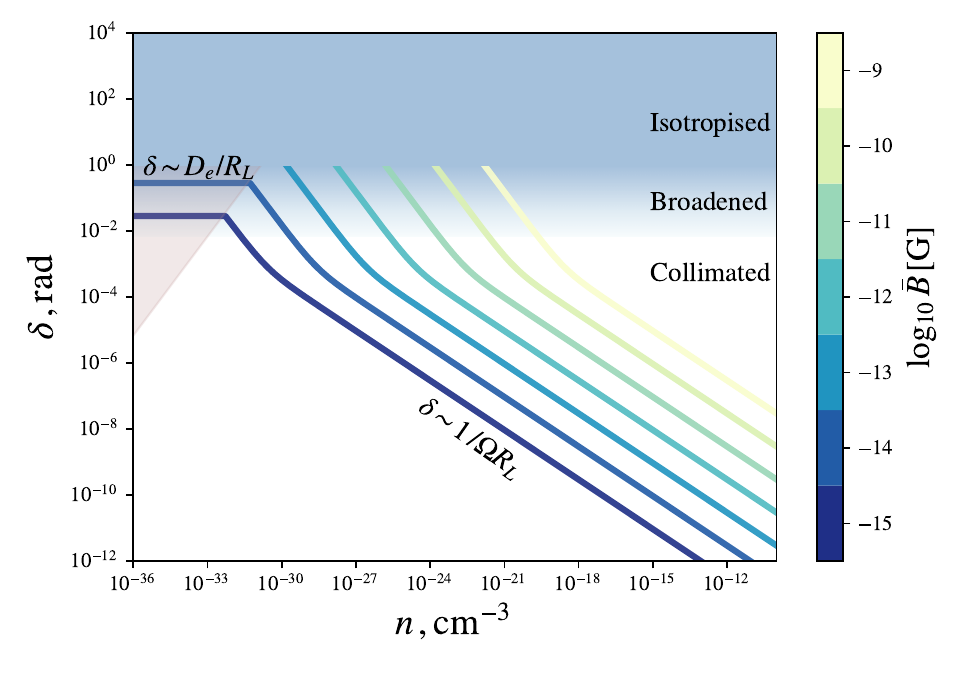}
    \caption{\label{fig:deflection-angle}
    Predicted deflection angle.  We plot the angle $\delta$ given by \eref{deflection-angle} 
    as a function of the number density of monopoles $n = \nM + \nMb$.  
    For this example we take 
the monopole magnetic charge to a single Dirac charge $\bar{g} = 2 \pi/ \bar{e}$, and the monopole mass to be $m = 100 \, \mathrm{GeV}$. 
    We further suppose the initial energy of the electron (or positron) to be $E_e = 3.5 \, \mathrm{TeV}$, so that it travels a distance $D_e \approx 100 \, \mathrm{kpc}$. 
We consider values for the amplitude of the (effectively homogeneous) IGMF in the range from $\bar{B} = 10^{-15}$ to $10^{-9} \, \mathrm{G}$.  
The pale red region 
    indicates where the monopole density is so low that it has no appreciable impact on $\delta$. 
    In this case, $\delta$ is given 
    by the first branch of \eref{deflection-angle}, with $\delta \sim D_e / R_L$.     
    Outside of this region, the monopole number density is high enough for the deflection angle to be affected by the monopole ``collimation'' effect. In this regime, the deflection angle follows the second branch of \eqref{deflection-angle}, $\delta \sim 1 / (\Omega R_L)$.
    The blue region (``Isotropised'') indicates where the deflection angle becomes as large as $\delta \sim 1$ so that the emission is effectively isotropized and contributes to the diffuse GeV background rather than being associated with any particular source. The unshaded region (``Collimated'') corresponds to deflection angles small enough such
that the angular size of the secondary halo from the blazar 1ES 0229+200 remains within the Fermi telescope’s point spread function (PSF), i.e., $\delta \lesssim \tau \theta_\text{psf}$ (with $\theta_\text{psf} \sim 0.2^\circ$ and $\tau \sim 2$). 
The region (``Broadened''), shown as a blue gradient, corresponds to deflection angles for which the secondary radiation is not fully isotropized, yet the angular size of the halo exceeds the PSF of the Fermi telescope. 
    }
\end{figure}

%=====
In \fref{fig:deflection-angle} we illustrate how the deflection angle $\delta$ varies with the monopole number density $n$,
focusing on monopoles with $g = \bar{g}$ and $m = 100 \, \mathrm{GeV}$.
For small $n$ the curves become approximately horizontal.  
In this regime, the low monopole density suppresses the magnetic field oscillation frequency~$\Omega$ and hence the deflection angle is given by the usual relation, $\delta \sim D_e / R_L$, which is independent of $n$.  
In other words, the time scale for magnetic Langmuir oscillations is much larger than the length of the electron trajectory, and the electrons ``see'' an effectively static magnetic field.

On the other hand with a sufficiently large~$n$ such that the oscillation frequency $\Omega$ exceeds $D_e^{-1}$, 
the deflection angle becomes $\delta \sim 1 / \Omega R_L$. 
In this regime, the deflection angle decreases with $n$.
This is because the magnetic Langmuir oscillations cause the effective magnetic field strength to be smaller, which leads to a smaller deflection.    
For $m = 100 \, \mathrm{GeV}$ the monopoles are light enough that they can be accelerated to relativistic speeds, which leads to $\delta \propto \Omega^{-1} \approx \Omega_0^{-1} \gmax^{1/2} \propto n^{-1}$ for intermediate values of $n$. For higher values of $n$, the monopoles remain nonrelativistic while they are being accelerated by the magnetic field, which leads to $\delta \propto \Omega^{-1} \approx \Omega_0^{-1} \propto n^{-1/2}$.

%====================
%  Monopoles abundance bounds
%====================
\section{Monopole abundance bounds}
\label{sec:monopoles-bounds}

%=====

In the previous section we described how an electron or positron is deflected by an angle $\delta$, given by \eref{deflection-angle}, when passing through a magnetic field.  
We argued that the effect of monopole-induced magnetic Langmuir oscillations is to reduce the deflection angle.  
In this section we discuss how the oscillations impact constraints on the IGMF from observations of TeV blazar halos {and can be used to constrain the monopole abundance, as summarized in the bottom row of fig.~\ref{fig:blazar-exeperiment-cartoon}.}

For our constraints, we adopt the parameters of blazar 1ES 0229+200 from the combined H.E.S.S. and Fermi-LAT spectrum analysis in \rref{Neronov_2010}.
The fiducial parameter values are given in Sec.~\ref{sec:blazar-fiducail}.

%--------------------
%  Monopole abundance bound
%--------------------
\subsection{Monopole abundance bound}
\label{sec:monopole_bound}

%=====

Recall from \sref{sec:blazar} that 
the deflection angle $\delta$ needs to satisfy the lower bound (\ref{eq:deltaLB}) to explain why blazar halos have thus far evaded detection. 
Here, since $\delta$ is a non-increasing function of the monopole number density~$n$, 
the lower bound (\ref{eq:deltaLB}) can be expressed as an upper bound on $n$ if an IGMF strength $\bar{B}$ is given.  

To derive the monopole bound,
we combine (\ref{eq:deltaLB}) with 
$\delta \lesssim 1/R_L \Omega$ which follows from (\ref{deflection-angle}), to obtain 
$\tau \theta_{\ro{psf}} \lesssim 1/R_L \Omega$.
We then substitute the monopole oscillation frequency from eq.~\eqref{magnetic-plasma-frequency}, the monopole Lorentz-factor amplitude from eq.~\eqref{energy-conserv}, and the electron/positron Larmor radius from eq.~\eqref{R-larmor} with $v_{e,\perp} \sim 1$. 
Combining these expressions yields an upper bound on the monopole number density:
\begin{equation}\label{eq:n_upper_bound} 
\begin{split}
   n &\lesssim \frac{\bar{B}^2}{\tilde{m}} 
\, f\left(\frac{m}{\tilde{m}}\right)
\\
& \approx \bigl( 2 \times 10^{-32} \, \mathrm{cm}^{-3} \bigr) \biggl( \frac{\bar{B}}{10^{-15} \, \mathrm{G}} \biggr)^{\!2} 
\left( \frac{\tilde{m}}{2 \, \ro{TeV}} \right)^{-1}
f\left(\frac{m}{\tilde{m}}\right) \, ,
\end{split}
\end{equation}
where $f(x) = x + \sqrt{1 + x^2}$, and 
\begin{equation}
    \tilde{m}  \equiv \sqrt{2}\, \tau \theta_{\mathrm{psf}}
    \frac{g E_e}{\bar{e}} 
    \approx 2 \, \mathrm{TeV}
    \left( \frac{ \tau \theta_{\mathrm{psf}}}{7\!\times\!10^{-3}} \right) \!
    \left(\frac{E_{\gamma, s}}{39\, \mathrm{GeV}} \right)^\frac{1}{2}\!
     \frac{g}{\bar{g}}\,.
\end{equation}
Here, upon moving to the far right-hand side we used \eref{E-gamma-IC}
to rewrite the electron/positron energy~$E_e$ in terms of the secondary photon energy $E_{\gamma,s}$; we set its reference value to the energy $E_{\gamma,\text{min}} \approx 39\, \ro{GeV}$ below which the EM cascade needs to be suppressed. 
The fiducial value $\tau \theta_{\mathrm{psf}} \sim 7 \times 10^{-3}$  corresponds to the product of 
$\theta_\text{psf} \sim 0.2^\circ$ and 
$\tau \sim 2$ chosen from the range $\tau \sim 1$--$4$ 
(cf. \sref{sec:blazar-fiducail}).

\begin{figure*}[t]
    \centering
    \includegraphics[width=\textwidth]{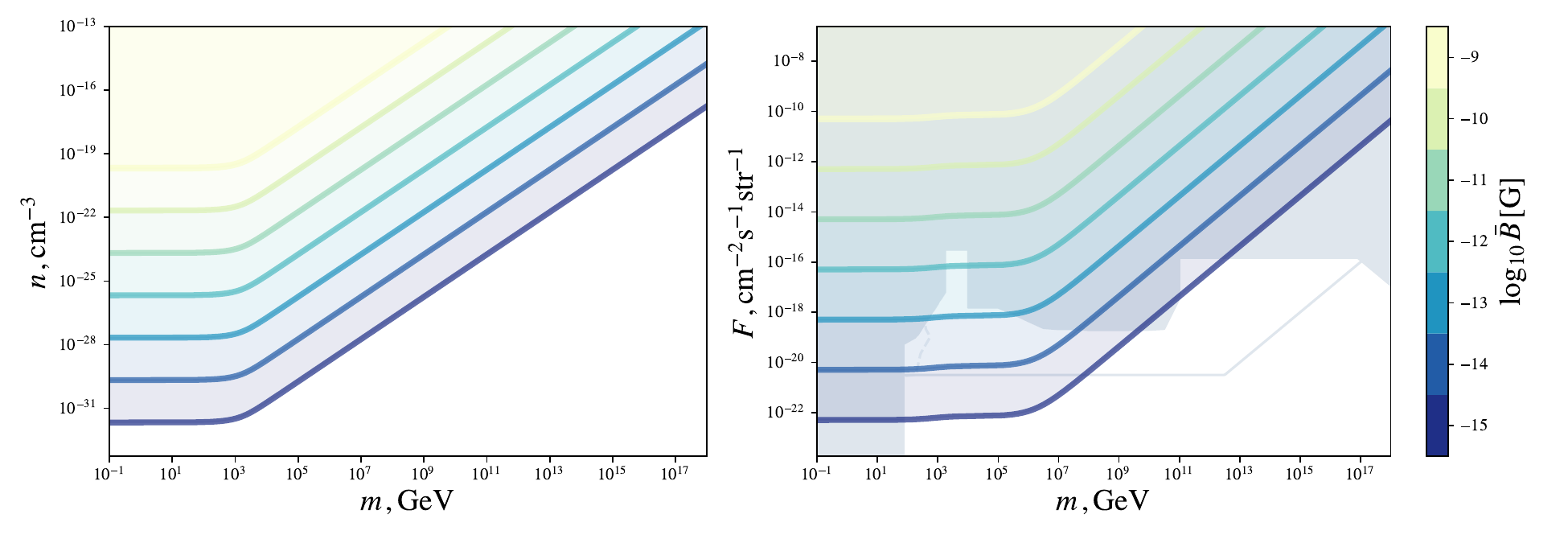}
    \caption{
    Monopole abundance bound.  We plot upper bounds on the monopole number density~$n$ in the universe given by \eref{eq:n_upper_bound}, and on the monopole flux~$F$ on Earth which is approximately given by eq.~(\ref{eq:FUB}).
We show values for the monopole mass $m $ ranging from $10^{-1}$ to $10^{18} \, \mathrm{GeV}$, and we show several values of IGMF strength $\bar{B}$ ranging from $10^{-15}$ to $10^{-9} \, \mathrm{G}$.  We fix the monopole magnetic charge to be a single Dirac charge $g = \bar{g}$, and the limits are derived from Fermi-LAT and H.E.S.S. observations of the blazar 1ES 0229+200, as described above \eref{eq:ext_vs_psf}. In the right panel, the gray shaded region represents a conservative combination of current constraints on monopole parameter space from MoEDAL \cite{Acharya_2022}, MACRO \cite{MACRO_2002}, IceCube~\cite{IceCube:2021eye}, Auger~\cite{PierreAuger:2016imq}, the galactic Parker bound \citep{Parker:1970xv,Turner_1982},
and the cosmic abundance bound,
with the experimental limits shown in terms of the monopole mass based on the analysis in~\cite{Perri_2025}.
The Auger bound can go down to the position of the dashed gray line depending on the IGMF strength. 
The solid gray line shows the seed Parker bound~\cite{Adams:1993fj} 
under the assumptions of a seed galactic magnetic field of $10^{-11}$~G and IGMF strength $\le 10^{-13}$~G \cite{Perri_2024}. Details on the existing constraints can be found in \aref{app:mono_constraints} and in \rref{Perri_2025}.
    }
    \label{fig:project_sensitivity}
\end{figure*}

In the left panel of \fref{fig:project_sensitivity} we present our upper bounds on the monopole number density $n$, inferred through the non-observation of GeV cascade emission from the blazar 1ES 0229+200.  
The bounds are plotted as functions of the monopole mass~$m$, 
with the different colors denoting different values of the IGMF strength~$\bar{B}$ as indicated in the legend. 
At $m \ll \tilde{m} \sim \ro{TeV}$ the function $f$ approaches unity and hence the upper bound is independent of $m$. 
At $m \gg \tilde{m}$, the function becomes $f \approx 2 m / \tilde{m}$ and so the bound weakens as $\propto m$. 
The bound also has an overall scaling 
with the IGMF strength of $\propto \bar{B}^2$. 

Note that with the maximum number density~$n_{\mathrm{UB}}$ allowed in (\ref{eq:n_upper_bound}), 
$m \ll \tilde{m}$ corresponds to $ m n_{\ro{UB}} \ll \bar{B}^2$, and vice versa.
Hence with the maximally allowed monopole abundance, 
$\tilde{m}$ sets the threshold mass below which the monopole oscillation is relativistic.
One can also see this explicitly by evaluating the oscillation amplitude of the monopole velocity at the maximum density,
\begin{equation}
 \left. \bar{v} \right|_{n = n_{\mathrm{UB}}}
\sim \min \left( 1, \, \frac{\bar{B}}{\sqrt{m n_{\mathrm{UB}}}}  \right)
\sim  \min \left(
1, \,  \frac{\tilde{m}}{\sqrt{2} m}
\right).
\label{eq:vmax_UL}
\end{equation} 
Here in the first equality we expressed the velocity in the relativistic and nonrelativistic regimes using (\ref{energy-conserv}),
and in the second equality we substituted (\ref{eq:n_upper_bound}) into $n_{\ro{UB}}$.

%--------------------
%  Fig 6
%--------------------

%=====

We can further rewrite the upper bound on the monopole density $n$ as an upper bound on the monopole flux $F$.  
We calculate the flux as 
\begin{equation}\label{n-to-flux}
    F = \frac{n  v_{\ro{MW}}}{4\pi} \,,
\end{equation}
where $v_{\ro{MW}}$ is the typical relative velocity of the monopoles and the Milky Way galaxy.  
Here, note that the oscillation amplitude of the monopole velocity~$\bar{v}$ represents the typical velocity of monopoles in the CMB rest frame, while the peculiar velocity of the Milky Way is of $v_p \sim 10^{-3}$. 
We take $v_{\ro{MW}}$ from the larger of the two velocities, by expressing it as 
\begin{equation}\label{n-to-flux:velocity}
    v_{\ro{MW}} \sim \bar{v} + v_p \, .
\end{equation}
Here we note that if the monopole abundance is sufficiently small such that the IGMF oscillation frequency is smaller than the Hubble rate, i.e. $\Omega < H_0$, then $\bar{v}$ in the above expression for $v_{\ro{MW}}$ should be replaced by the velocity the monopoles acquire within the age of the universe (see app.~\ref{sec:vsHubble}). 
However, the blazar observations constrain monopoles only if they give rise to magnetic field oscillations with $\Omega > D_e^{-1} \gg H_0$ (cf. (\ref{deflection-angle})).
Hence for our purpose of obtaining upper bounds on the monopole flux from blazar observations, the expression (\ref{n-to-flux:velocity}) suffices.

We also remark that $F$ in (\ref{n-to-flux})
denotes the incident monopole flux on the Milky Way, which is equivalent to the flux inside the Milky Way 
(even if the monopoles are further accelerated by galactic magnetic fields)
from the conservation of the number of monopoles.\footnote{This is not the case if the monopoles are captured by the Milky Way's gravitational potential. However monopoles with a Dirac charge stay clustered with the Milky Way until today only if they are ultraheavy as $m \gtrsim 10^{18}\, \ro{GeV}$ \cite{Turner_1982,Kobayashi_2023}.}
Moreover, since the velocity of Earth with respect to the Milky Way is also of $10^{-3}$, 
$F$ also gives the monopole flux on Earth. 

The upper bound on magnetic monopoles flux is shown in the right panel of \fref{fig:project_sensitivity}. 
In order to understand its mass dependence, let us 
also derive an analytic expression for the flux bound by approximating the number density bound in (\ref{eq:n_upper_bound}) as $n_{\mathrm{UB}} \sim (\bar{B}^2 / \tilde{m}) \max (1, 2 m / \tilde{m}) $, the relative velocity in (\ref{n-to-flux:velocity}) as
$v_{\ro{MW}} \sim  \max (\bar{v}, \, v_p)$, 
and substituting them together with (\ref{eq:vmax_UL}) into the flux~(\ref{n-to-flux}).\footnote{The flux scales with the number density as $F \propto \sqrt{n}$ (if $v_{\ro{MW}} \sim \bar{v} \ll 1$) or
$F \propto n$ (otherwise).}
One can then check that the combinations of the $\max$ and $\min$ functions yield two distinct behaviors for the mass dependence of the flux bound as
\begin{widetext}
\begin{equation}\label{eq:FUB}
\begin{split}
  F &\lesssim  \frac{ \bar{B}^2}{4\pi \tilde{m}} \, 
\mathrm{max} \left( 1, \, \frac{ 2v_{p} m }{\tilde{m}} \right) \\
&\approx (6 \times 10^{-23}\, \mathrm{cm}^{-2} \mathrm{sec}^{-1} \mathrm{str}^{-1}) 
\left( \frac{\bar{B}}{10^{-15}\, \mathrm{G}} \right)^2
\left( \frac{\tilde{m}}{2\, \mathrm{TeV}} \right)^{-1}
\mathrm{max}\left\{ 1, \, 
\left(\frac{v_p}{10^{-3}}\right)
\left( \frac{\tilde{m}}{2 \, \ro{TeV} }  \right)^{-1}
\frac{m }{10^3\, \ro{TeV}} 
\right\} 
\,.
\end{split}
\end{equation}
\end{widetext}
%\mariia{$8 \to 5$}
Below the mass threshold 
$\tilde{m }/ 2 v_p \sim 10^3\, \mathrm{TeV}$,
the velocity of monopoles with the maximally allowed abundance 
$ \left. \bar{v} \right|_{n = n_{\mathrm{UB}}}$
is larger than $v_p$, and hence the relative velocity takes
$ \left. v_{\ro{MW}} \right|_{n = n_{\mathrm{UB}}} \sim 
\left. \bar{v} \right|_{n = n_{\mathrm{UB}}}$. 
In this regime, 
$ \left. \bar{v} \right|_{n = n_{\mathrm{UB}}}$ takes either
relativistic or nonrelativistic values 
at $m \ll \tilde{m}$ and 
$\tilde{m} \ll m \ll \tilde{m} / v_p $, 
respectively;
however in both cases, 
the $m$~dependencies of 
$n_{\ro{UB}}$ (cf. (\ref{eq:n_upper_bound})) and
$\left. \bar{v} \right|_{n = n_{\ro{UB}}}$ (cf. (\ref{eq:vmax_UL}))
cancel out from their product. 
As a consequence, the flux bound is nearly $m$~independent at $ m \ll \tilde{m} / v_p$.\footnote{The product $(n v_{\ro{MW}})_{n = n_{\ro{UB}}}$ at $m \ll \tilde{m}$ and $\tilde{m} \ll m \ll \tilde{m} / v_p $
actually differ by a factor~$\sqrt{2}$. This is seen in the plot, however we ignored this in the approximate expression (\ref{eq:FUB}).} 
At $ m \gg \tilde{m} / v_p$, the bound weakens as $\propto m$.
The overall scaling of the bound $\propto \bar{B}^2$ with the IGMF strength is similar to the number density bound.

%=====
It is illuminating to compare the upper bound on $F$ that we have derived here against upper bounds on $F$ that were derived in previous work. 
Existing bounds appear as gray shaded regions in the right panel of \fref{fig:project_sensitivity}, and \aref{app:mono_constraints} discusses them in detail. 
The constraints on monopole flux for low-mass monopoles $m \lesssim 10^{10}$~GeV, derived in this work from the blazar–IGMF experiment, are stronger than the existing conservative bounds
by up to six orders of magnitude in the most optimistic case of $\bar{B} \sim 10^{-15}$~G. In the latter case, the constraint obtained in this work also exceeds 
the seed Parker bound~\cite{Adams:1993fj} shown by the solid gray line, which itself depends on assumptions about the seed galactic magnetic field and the IGMF strengths. At the same time, if the IGMF value is larger ($\bar{B} \gtrsim 10^{-12}$~G), our bounds become subdominant to existing constraints. Independent probes of the IGMF strength will help reduce the uncertainty in our bounds on monopole abundance.

%--------------------
%  IGMF strength bound
%--------------------
\subsection{IGMF strength bound}
\label{sec:IGMF_bound}

The deflection angle $\delta$ is a non-decreasing function of the IGMF field strength~$\bar{B}$, 
and thus the lower bound (\ref{eq:deltaLB}) on $\delta$ 
can also be expressed as a lower bound on $\bar{B}$ if a monopole number density $n$ is given.  
The expression (\ref{deflection-angle}) for $\delta$ implies both
$\delta \lesssim D_e / R_L$
and
$\delta \lesssim 1/R_L \Omega$.
Combining these with (\ref{eq:deltaLB}) and proceeding similarly for the bound on $n$, we obtain
\begin{widetext}
\begin{equation}\label{eq:B0_upper_bound}
   \bar{B} \gtrsim \mathrm{max} 
   \begin{dcases}
   \tau \theta_\mathrm{psf} \frac{E_e}{\bar{e} D_e}  \\ 
   (n \tilde{m})^{1/2}
 \left[  f \left(\frac{m}{\tilde{m}}\right) \right]^{-1/2}
   \end{dcases} 
\approx
\max
\begin{dcases}
    \bigl( 2 \times 10^{-16} \, \mathrm{G} \bigr)
    \left( \frac{ \tau \theta_{\mathrm{psf}}}{7\!\times\!10^{-3}} \right)
\left(\frac{E_{\gamma, s}}{39\, \mathrm{GeV}} \right)
 \\ 
    \bigl( 6 \times 10^{-15} \, \mathrm{G} \bigr) \Bigl( \frac{n}{10^{-30} \, \mathrm{cm}^{-3}} \Bigr)^{\!1/2} 
\left( \frac{\tilde{m}}{2 \, \ro{TeV}} \right)^{1/2}
 \left[  f \left(\frac{m}{\tilde{m}}\right) \right]^{-1/2}
    \end{dcases}
   \,.
\end{equation}
\end{widetext}
The first line derives from $\delta \lesssim D_e / R_L$, and 
corresponds to the standard lower bound on the IGMF strength from blazar observations~\cite{Neronov_2010}, independent of the presence of monopoles.  
We have treated the magnetic field as effectively homogeneous for the electrons and positrons, hence this bound is applicable if the IGMF coherence length $\lambda_B$ is large compared to $D_e \sim 100 \, \mathrm{kpc}$; the bound improves for $\lambda_B \lesssim D_e$~\cite{Neronov_2010}.  
The second line is from $\delta \lesssim 1/R_L \Omega$,
and so is equivalent to \eref{eq:n_upper_bound}.   
This is a new lower bound on $\bar{B}$, which becomes stronger than the usual bound if the monopole density is sufficiently high and the magnetic field is oscillating sufficiently rapidly.  

\begin{figure}[h!]
    \centering
    \includegraphics[width=\linewidth]{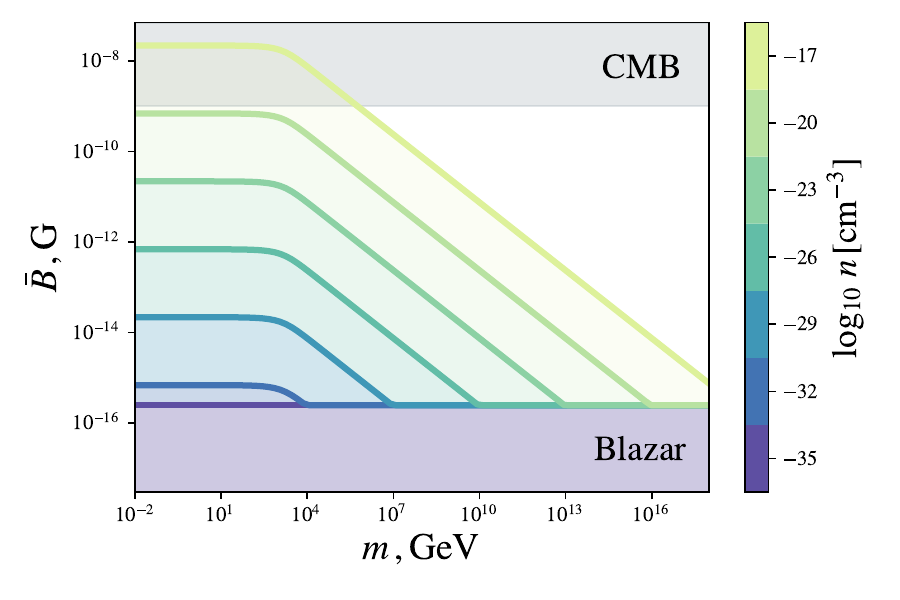}
    \caption{
    IGMF strength bound.  We plot lower bounds on the strength of the IGMF $\bar{B}$ given by \eref{eq:B0_upper_bound}. 
    We show a range of values for the monopole mass $m $ ranging from $10^{-2}$ to $10^{17} \, \mathrm{GeV}$, and we show several values of   the monopole number density $n$ ranging from $10^{-35}$ to $10^{-17} \, \mathrm{cm}^{-3}$.  
    We fix the monopole magnetic charge to a single Dirac charge $g = \bar{g}$, and the limits are derived from Fermi-LAT and H.E.S.S. observations of the blazar 1ES 0229+200, as described above \eref{eq:ext_vs_psf}.  The limits apply for IGMFs with coherence lengths larger than $\lambda_B > D_e \sim 100 \, \mathrm{kpc}$.  The gray shaded region indicates an upper limit on the IGMF strength derived from observations of the CMB spectrum and anisotropies~\cite{Planck:2015zrl,Zucca:2016iur,Jedamzik:2018itu}. 
    }
    \label{fig:magnetic-field-constraints}
\end{figure}

%=====
In \fref{fig:magnetic-field-constraints} we present our lower bound on the strength of the IGMF, inferred through the non-observation of GeV cascade emission from the blazar 1ES 0229+200.  
For small $n$, the monopole density is low, and the monopole-induced magnetic Langmuir oscillations are slow, so the lower bound on this effectively static magnetic field $\bar{B}$ reduces to the usual one~\cite{Neronov_2010}.  
For larger $n$, the monopole density is higher, and the magnetic Langmuir oscillations lead to reduced broadening of the electromagnetic cascade, since the magnetic field is oscillating rapidly in comparison with the time scale for the electron/positron trajectories.  
Consequently, a stronger IGMF strength (larger $\bar{B}$) is required to explain the non-observation of blazar cascade emission.  
If the monopoles are heavier (larger $m$), then their oscillations are slower [cf., (\ref{magnetic-plasma-frequency})], and the lower bound on $\bar{B}$ becomes weaker.   
For comparison we also show an upper limit on the strength of cosmological magnetic fields that has been derived from observations of the CMB radiation, but we caution that the CMB bound may also be modified if the time scale for magnetic Langmuir oscillations is shorter than the duration of recombination.  

%--------------------
%  Fig 7
%--------------------

%====================
%  Discussion
%====================
\section{Critical assessment of assumptions}
\label{sec:discussion}

%=====

In order to establish the robustness of our results, in this section we assess the validity of the assumptions made in the main text and summarize the calculations presented in several appendices.
The constraints on monopole abundance presented in this work 
rely on the assumption that an IGMF exists and plays the primary role
in explaining the non-observation of blazar secondary GeV photon halos by deflecting charged particles within the EM cascade.
This, however, is not the only possible explanation. An alternative hypothesis for the non-observation of blazar secondary photon halos is the development of beam-plasma instabilities within the EM cascade, providing a mechanism for electron--positron pair energy dissipation without contributing to secondary GeV photon emission~\citep{Broderick_2012, Broderick_2014, Chang_2014, Chang_2016, Shalaby_2018, Broderick_2018, Shalaby_2020, Lamberts_2022}.
(See also the recent work \cite{Arrowsmith:2025apl} which questioned this hypothesis.)
Several studies have examined the IGMF interpretation of TeV blazar observations. Here, we summarize some of these results. Ref.~\cite{Blanco:2023kfa} studied the contribution of blazars to the isotropic $\gamma$-ray background and claimed that it appears to be in tension with observations.
\Rref{Arlen:2012iy} argues that, when various uncertainties in source modeling are taken into account, such as the poorly known blazar duty cycle, blazar observations are consistent with zero IGMF.
Other studies~\cite{Essey:2010nd,Essey:2010er} explore the assumption that secondary photons produced along the line of sight by cosmic-ray interactions with EBL or CMB photons dominate the signal, and conclude that this scenario is consistent with IGMF lower limit of $10^{-17}$~G.

%=====
The constraints on magnetic monopole abundance inferred in this work are also affected by various sources of uncertainty inherited from the blazar experiment.
Variations in the EBL model lead to an order-of-magnitude uncertainty in the optical depth $\tau$ to the source~\cite[see, e.g.,][]{Neronov_2009}. 
The value of $E_{\gamma, \text{min}}$, which also directly affects the monopole bounds, depends on the modeling of the Fermi-LAT and H.E.S.S. spectra of the blazar, and can be influenced by the degeneracy between the power-law index and the cut-off energy.
Also, the Fermi-LAT and H.E.S.S. observations of blazar 1ES 0229+200 were conducted several years apart, and therefore do not account for possible time variation of the source~\cite{Neronov_2010}. 

In our main analysis, we modeled the monopole population as a continuum, and used density and bulk velocity fields to describe the monopole-magnetic field oscillation. This treatment is justified
in app.~\ref{app:C2};
there we show that the continuum treatment breaks down only 
if the monopole density takes very large values that are
already ruled out by existing experiments
(see discussions around (\ref{eq:31})).

We have also assumed that the IGMF is effectively homogeneous
so that electric fields sourced by spatial gradients of magnetic fields are negligible, and also that the magnetic field oscillates coherently over the distance~$D_e$ traveled by the electrons and positrons in the cascade. 
We show in app.~\ref{sec:inhomo} that 
the homogeneous approximation is valid
as long as the IGMF correlation length satisfies $\lambda_B \gg D_e $.

%=====

Another question regarding the monopole-induced oscillations of the magnetic field is whether the coherent oscillations can be preserved for a sufficiently long time.
Landau damping is shown to be negligible in app.~\ref{sec:inhomo}.
Instead, the main dissipation mechanisms in this system are inverse Compton (IC) radiative losses by ultra-relativistic monopoles accelerated by the IGMF, 
and energy losses via Coulomb scattering with electrons in the intergalactic plasma (see app.~\ref{app:dissipation}). These mechanisms are most efficient for low-mass monopoles, $m \lesssim 10^3$~GeV.
For these low-mass monopoles, within a certain range of number densities, the energy dissipation rate is sufficiently high that the IGMF energy losses become comparable to its total energy density, potentially leading to significant IGMF depletion; see~\fref{fig:ic-energy-loses} and \fref{fig:coulomb-dissipation}. Experiments such as MoEDAL can probe monopole masses in the range where monopole-induced dissipation would be incompatible with IGMF survival.

Our blazar limits on monopoles can be modified if the IC and Coulomb dissipation efficiently damps the magnetic field oscillation. 
However, parameter regions which thus may evade the blazar limit 
exist only at very small masses already excluded by MoEDAL
(see Fig.~\ref{fig:bounds-vs-dissipation}).
Overall, dissipation effects may lead to interesting phenomenology for monopoles with number densities or fluxes lower than those probed by blazar observations discussed in this work.

Other considered dissipation mechanisms, such as Larmor radiation, are negligible for monopoles with fluxes not already excluded by existing bounds (see \aref{app:dissipation}).
At ultra-high energies, monopoles may produce particle pairs when scattering off photons or particles in the IGM through processes such as Bethe–Heitler electron–positron pair production. However, these channels are expected to be subdominant compared to the dissipation mechanisms discussed above and are therefore not considered further here.

%=====
At the same time, scatterings with the plasma or CMB contribute not only to dissipation but also to increasing the velocity dispersion of monopoles.
We studied the effect of an arbitrary monopole velocity distribution on monopole–magnetic field oscillations (see \aref{app:velocity-dsitribution}), and 
showed that under rather generic conditions,
the magnetic field oscillation profile and frequency remain unaffected.
External non-EM forces also do not affect the magnetic field oscillations, provided that such forces do not accelerate monopoles to relativistic Lorentz factors in the absence of a magnetic field (see \aref{app:external-forces}), which remains a valid assumption for gravitational forces far from massive structures.  Altogether, while theoretical and observational uncertainties remain, our monopole constraints provide a complementary probe to existing laboratory and astrophysical bounds.

%====================
%  Conclusions
%====================
\section{Conclusions}
\label{sec:conclusions}

%=====
Observations of TeV blazars provide strong evidence for the existence of an IGMF with $B \gtrsim 10^{-15}~\mathrm{G}$. An IGMF of that strength can explain the non-observation of GeV secondary photons in the blazar spectrum due to the deflection of charged particles within the EM cascade, which redistributes secondary $~$GeV photons out of the line of sight.

%=====
The presence of magnetic monopoles in the intergalactic space causes a magnetic analogue of Langmuir oscillations. Monopole-induced magnetic field oscillations modify the deflection angle of electrically charged particles, and when the monopole abundance is high enough, the oscillation frequency~$\Omega$ exceeds the Larmor frequency~$\omega_c$ of the particles, leading to a ``collimation'' effect that reduces the deflection angle as $\propto \omega_c / \Omega$.
This results in a smaller angular size of the secondary photon halo in blazars. Therefore, the non-observation of GeV photons from EM cascades of blazars can be used to probe the monopole abundance.

%=====
We derived a constraint on monopole abundance from the H.E.S.S. and Fermi-LAT limits on the secondary halo angular size for the blazar 1ES 0229+200. 
The upper bound on the monopole flux is presented in Fig.~\ref{fig:project_sensitivity} and Eq.~(\ref{eq:FUB}). 
Our constraint depends on the IGMF strength, whose exact value is still unknown. 
The bound on low-mass monopoles with $m \lesssim 10^6\, \text{GeV}$, in the most optimistic scenario of IGMF $B \sim 10^{-15}$~G, reaches down to $F \lesssim 6 \times 10^{-23} \text{cm}^{-2} \text{s}^{-1} \text{str}^{-1}$.
This is stronger than current conservative bounds by up to six orders of magnitude; it is also an order of magnitude stronger than the seed Parker bound, which also depends on the IGMF and seed galactic magnetic field strengths. 
The bound remains stronger than current conservative constraints for IGMF up to $B \lesssim 10^{-12}$~G and becomes subdominant if the IGMF value is higher than that. Independent probes of the IGMF strength would help reduce the uncertainty in our magnetic monopole abundance constraints.

%=====
We also revisited the IGMF constraints from TeV blazar observations assuming a non-zero magnetic monopole abundance, and showed that these lower bounds on the IGMF become stronger for higher monopole abundances and lower monopole masses. 

%=====
We remark that throughout this paper, we assumed the IGMF to have a correlation length of $\lambda_B \gg 100\, \ro{kpc}$. 
It would be interesting to analyze monopole bounds for smaller correlation lengths, although understanding the 
plasma oscillation in such cases  will likely require numerical simulations. We leave an investigation of this for the future.

%====================
%  Acknowledgements
%====================
\section*{Acknowledgements}

We thank Mustafa Amin, Carlos Blanco, Stephen Bradshaw, Axel Brandenburg, Michele Doro, Oksana Iarygina, Mariangela Lisanti, David Marsh, Daniele Perri,  Anirudh Prabhu, Sandip Roy, Piero Ullio, Tanmay Vachaspati, and Matteo Viel for valuable discussions and insights. 
This material is based upon work supported (in part: A.J.L.) by the National Science Foundation under Grant No.~PHY-2412797. 
The works of M.K. and T.K. were supported by the European Union - NextGenerationEU through the PRIN Project ``Charting unexplored avenues in Dark Matter'' (20224JR28W), and by INFN TAsP.
T.K. also acknowledges support from JSPS KAKENHI (JP22K03595 and JP26K07064).

\appendix

%====================
%  Summary of magnetic monopole constraints
%====================
\section{Summary of magnetic monopole constraints}
\label{app:mono_constraints}

%=====
In this appendix we summarize various astrophysical and laboratory constraints on the existence and abundance of magnetic monopoles.  
Additional discussion is available in \rrefs{ParticleDataGroup:2024cfk,Perri_2025}.
Here, unless otherwise stated,
we consider monopoles with a single Dirac charge, i.e. $g = \bar{g}$.

%===== 
The existence of magnetic monopoles is tested by the MoEDAL experiment at the LHC.  
For monopoles with charge $g/\bar{g} = 1$, $2$, or $3$, the MoEDAL experiment obtained a lower bound on the mass $m > 75 \, \mathrm{GeV}$ at $95\%~\mathrm{C.L.}$, through the non-observation of monopole / anti-monopole pairs produced via the Schwinger effect \cite{Acharya_2022}.  
Other mass lower limits have also been obtained by assuming Drell--Yan and photon-fusion monopole production~\cite{Acharya_2025};
however for such processes, there are considerable theoretical uncertainties in the production cross section, and moreover, they are expected to be strongly suppressed for solitonic monopoles due to their high degree of compositeness~\cite{Witten:1979kh,Drukier:1981fq}.
We hence do not show those limits in Fig.~\ref{fig:project_sensitivity}.\footnote{We also do not show constraints from catalysis of nucleon decay, since the catalysis depends on the details of the monopole model.}

%=====
Rather than creating monopoles in the laboratory, other experiments search for an existing population of monopoles.  
The MACRO experiment uses liquid scintillator detectors to search for the tracks that would be induced by the transit of a magnetic monopole and to measure their velocity.  
The non-observation of these tracks furnished constraints on the monopole flux for a range of monopole masses and velocities.   
We adopt the bound $F \le 1.4 \times 10^{-16}~\text{cm}^{-2}~\text{s}^{-1}~\text{str}^{-1}$~\cite{MACRO_2002} for monopoles with masses above $m \gtrsim 10^{10}$~GeV, following the monopole velocity arguments from~\cite{Perri_2025}.

%=====
Telescopes that search for ultra-high energy cosmic rays are also equipped to probe a local relativistic population of magnetic monopoles.  
The IceCube neutrino observatory constrains the flux down to 
$F\lesssim 10^{-19}~\text{cm}^{-2}~\text{s}^{-1}~\text{str}^{-1}$ \cite{IceCube:2021eye}
in the mass range from $10^4$ to $10^{11} \, \mathrm{GeV}$,
while the Pierre Auger cosmic ray observatory constrains
low mass monopoles $ m \lesssim 10^3 \, \mathrm{GeV}$
with the flux bound going down to
$F\lesssim 10^{-21}\,\text{cm}^{-2}~\text{s}^{-1}~\text{str}^{-1}$ \cite{PierreAuger:2016imq}
(see \cite{Perri_2025} for the derivation of the applicable mass ranges for both limits.)
The detailed mass dependence of the Auger limit depends on the IGMF strength, as the IGMFs affect the monopole velocity;
the Auger limit within the gray region in Fig.~\ref{fig:project_sensitivity} assumes $\bar{B} \lesssim 10^{-10}\, \ro{G}$, while for
$\bar{B} = 10^{-9}\, \ro{G}$ the limit goes down to the position of the gray dashed line~\cite{Perri_2025}.

%=====
Astrophysical considerations also lead to constraints on the flux of magnetic monopoles in our galaxy.  
Since magnetic monopoles would tend to deplete the strength of the galactic magnetic field, the survival of the Milky Way's micro-Gauss magnetic field implies an upper bound on the flux of magnetic monopoles.  
This limit, known as the Parker bound \cite{Parker:1970xv,Turner_1982}, 
gives $F \lesssim 3 \times 10^{-16}~\text{cm}^{-2}~\text{s}^{-1}~\text{str}^{-1}$ for monopoles with $m \lesssim 10^{17}$~GeV, and weakens as $\propto m$ above that mass 
(see also \cite{Kobayashi_2023,Long:2015cza}).
Based on a similar idea, a stronger bound of
$F \lesssim 3\times 10^{-21}~\text{cm}^{-2}~\text{s}^{-1}~\text{str}^{-1}$ follows from the survival of a seed galactic magnetic field assumed to be of $10^{-11}$~G~\cite{Adams:1993fj};
this bound is shown in Fig.~\ref{fig:project_sensitivity}
by the gray solid line. 
However, this seed Parker bound depends sensitively on the monopole acceleration in IGMFs, and the displayed bound holds if the IGMF strength is 
$\bar{B} \lesssim 10^{-13}$~G, while it becomes weaker for larger~$\bar{B}$~\cite{Perri_2024}.
Lastly, the flux bound at $m \gtrsim 10^{17}\, \ro{GeV}$ is dominated by the requirement that the monopole abundance should not exceed the dark matter density in the universe.

%====================
%  Monopole oscillation period
%====================
\section{Monopole oscillation period}
\label{app:period}

\begin{figure}[t]
    \centering
    \includegraphics[width=0.9\linewidth]{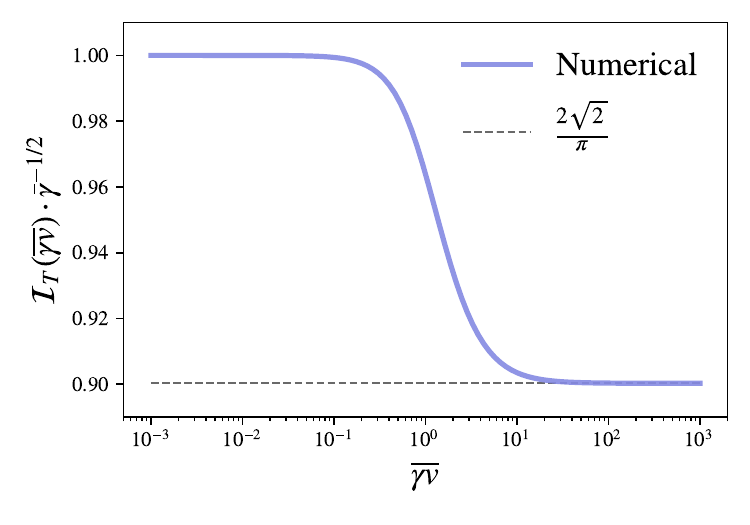}
    \caption{The numerical integral $\mathcal{I}_T(\overline{\gamma v})$. We show $\overline{\gamma v}$ in the range from $10^{-3}$ to $10^3$. The function $\mathcal{I}_T(\overline{\gamma v})$ is defined in \eqref{num-integral-period} and is divided by $\gmax^{1/2}$.
    The residual factor $\mathcal{I}_T(\overline{\gamma v}) / \gmax^{1/2}$ asymptotically approaches $1$ in the limit $\overline{\gamma v} \to 0$ and $2\sqrt{2}/\pi$ in the limit $\overline{\gamma v} \to \infty$.}
    \label{fig:num_I_T}
\end{figure}

In this appendix, we provide the detailed derivation of the period of magnetic field–monopole oscillations, for the setup presented in \sref{sec:magn-oscillations}. A quarter period of oscillations 
during which the Lorentz factor of the monopoles increases from $1$ to $\bar{\gamma}$ is
\begin{equation}
    \frac{T}{4} = \int\limits_{0}^{T/4} \mathrm{d}t 
= \int\limits_{1}^{\gmax} \frac{\mathrm{d} t}{\mathrm{d}\gamma} \mathrm{d}\gamma  \,.
\end{equation}
Noting that \eref{mm-eom} can be rewritten as
\begin{equation}
 \frac{\mathrm{d} \gamma }{\mathrm{d} t} = \frac{g B v}{m},
\end{equation}
and further rewriting $B$ in terms of $\gamma$ using \eref{energy-conserv}, we find:
\begin{align}
    T 
= \frac{2\pi}{\Omega_0} \frac{\sqrt{2}}{\pi} \int\limits_{1}^{\gmax} \frac{\gamma \mathrm{d}\gamma}{\sqrt{(\gamma^2 - 1)(\gmax - \gamma)}}  \equiv \frac{2\pi}{\Omega_0} 
    \mathcal{I}_T(\overline{\gamma v})\,.
\end{align}
In the final expression, we rewrote the integral in terms of 
$\overline{\gamma v }= \sqrt{\bar{\gamma}^2 - 1}$, so that 
\begin{equation}
    \mathcal{I}_T(x) = \frac{\sqrt{2}}{\pi} \int\limits_{0}^{x} 
    \frac{\mathrm{d}x'}{ \left\{ (1+x^2)^{1/2} - (1+x'^2)^{1/2} \right\}^{1/2}}\,.
    \label{num-integral-period}
\end{equation}
This function can be evaluated numerically and is approximately equal to $\mathcal{I}_T(\overline{\gamma v}) \approx (1 + (\overline{\gamma v})^2)^{1/4} = \gmax^{1/2}$, up to a factor of order unity. This numerical factor, as a function of $\overline{\gamma v}$, is shown in \fref{fig:num_I_T}.
In the nonrelativistic limit $\overline{\gamma v} \to 0$,
the integral $\mathcal{I_T} (\overline{\gamma v})$  approaches~$1$,
and so the period reduces to the harmonic value
$T = 2 \pi / \Omega_0$ as discussed below (\ref{b-harmonic}).
On the other hand in the ultrarelativistic limit 
$\overline{\gamma v} \to \infty$,
then $\mathcal{I_T} (\overline{\gamma v}) $ asymptotes to 
$ (2 \sqrt{2} / \pi) \sqrt{\bar{\gamma}}$,
and hence $T = 4 \sqrt{2 \bar{\gamma}} / \Omega_0$
as discussed below (\ref{b-triangular}). 

Neglecting order-unity coefficients, we can approximate the oscillation period and corresponding frequency for general values of $\overline{\gamma v}$ as
\begin{equation}
    T = \frac{2\pi}{\Omega} \,, \qquad 
    \Omega \approx \frac{\Omega_0}{ \bar{\gamma}^{1/2}} \,.
    \label{generalised-oscillation-period}
\end{equation}
These approximate expressions are used throughout the main text to model monopole–magnetic field oscillations.
We stress that the expression (\ref{generalised-oscillation-period}) becomes exact in the nonrelativistic limit, while in the ultrarelativistic limit it differs from the exact result only by a numerical factor of $\frac{\pi}{2\sqrt{2}} \approx 1.1$.

%--------------------
%  Fig 9
%--------------------

Let us also evaluate the spatial oscillation amplitude of the individual monopoles. This can be computed as the distance each monopole travels during the quarter period,
\begin{equation}
 d = \int_0^{T/4} \abs{v}\, \mathrm{d}t
 =  \int_0^{\bar{B}} \frac{\mathrm{d}B}{g n}
 = \frac{\bar{B}}{g n},
\label{eq:d}
\end{equation}
where in the second equality we used (\ref{maxwell}).

%====================
%  Monopoles Oscillations: Physical Scales
%====================

\section{Monopole oscillations: physical scales}

In this appendix, we provide information about the physical scales associated with monopole–magnetic field oscillations.  

\subsection{Plasma oscillation frequency and Hubble rate}
\label{sec:vsHubble}

Upon deriving blazar limits on the monopole density in the main text, we were interested in densities large enough to source plasma oscillation periods shorter than the distance traveled by the electrons/positrons~$D_e$, which in turn is much shorter than the Hubble radius~$H_0^{-1}$. 
Here we discuss the velocity of monopoles with a wide range of densities, including cases where the plasma oscillation does not finish one cycle within a Hubble time.

\begin{figure*}[t]
    \centering
    \includegraphics[width=\linewidth]{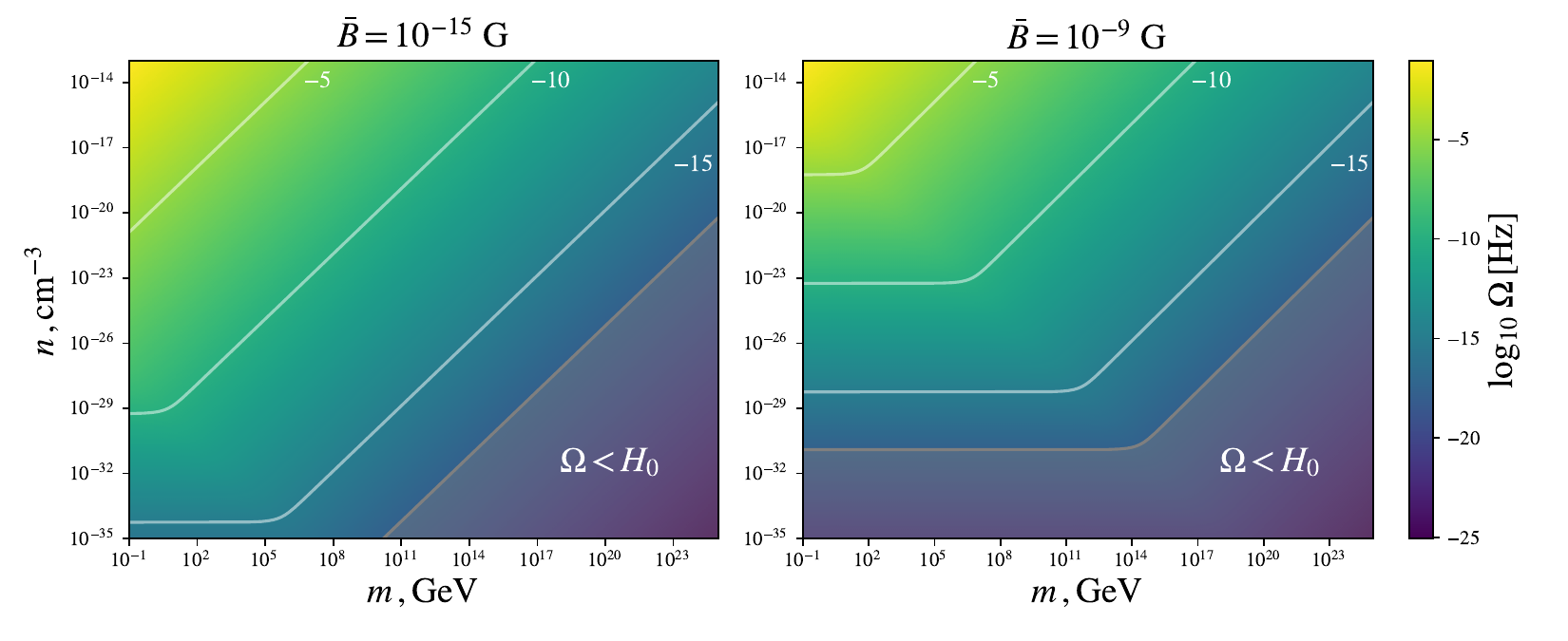}
    \caption{Monopole plasma oscillation frequency. The frequency for monopoles with mass $m$ and number density $n$ is shown in color.
The monopole charge is fixed to a single Dirac charge, while
the IGMF amplitude is fixed at $\bar{B} = 10^{-15}$~G (\textit{\bfseries left panel}) and $\bar{B} = 10^{-9}$~G (\textit{\bfseries right panel}). The gray shaded region marks where the oscillation frequency falls below the Hubble expansion rate.
}
    \label{fig:oscillation_freq}
\end{figure*}

The plasma oscillation frequency of the magnetic field–monopole system is given by~\eqref{generalised-oscillation-period}. Using \eqref{energy-conserv}, this becomes:
\begin{equation}
    \Omega \sim \sqrt{\frac{2g^2n^2}{\bar{B}^2 +2mn}}\,.
\end{equation}
In the two limiting regimes, the expression takes the following approximate forms.
In the ultra-relativistic case ($m n \ll \bar{B}^2$), the frequency becomes
\begin{equation}
    \Omega \sim  2 \cdot 10^{-11} \frac{g}{\bar{g}} \left(\frac{n}{10^{-30} \text{cm}^{-3}}\right) \left(\frac{\bar{B}}{10^{-15}\text{G}} \right)^{-1} \, \text{Hz}\,,
\label{eq:C2}
\end{equation}
while in the non-relativistic case ($m n \gg \bar{B}^2$):
\begin{equation}
    \Omega \sim  9 \cdot 10^{-16}\frac{g}{\bar{g}} \left(\frac{n}{10^{-30} \text{cm}^{-3}} \right)^\frac{1}{2} \left(\frac{m}{10^{10} \text{GeV}} \right)^{-\frac{1}{2}} \,\text{Hz}\,.
\end{equation}
In \fref{fig:oscillation_freq}, we plot $\Omega$ as a function of monopole mass and number density, for fixed IGMF amplitudes. 
As one moves towards smaller~$m$ the oscillation becomes relativistic and $\Omega$ approaches the value in (\ref{eq:C2}). 
On the other hand at sufficiently small densities,
$\Omega$ becomes smaller than the Hubble rate $H_0 \sim 2 \times 10^{-18}$~Hz. 

%--------------------
%  Fig 10
%--------------------

%--------------------
%  Fig 11
%--------------------
\begin{figure*}[t]
    \centering
    \includegraphics[width=\linewidth]{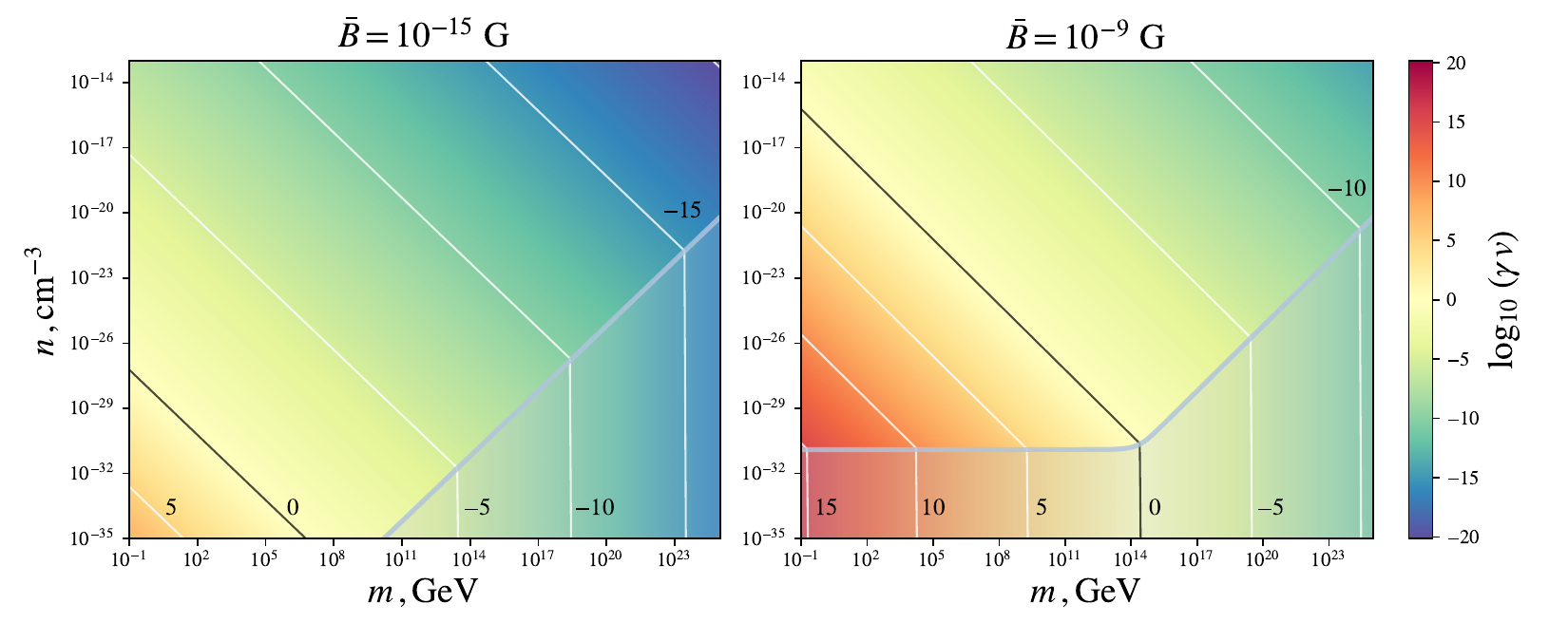}
    \caption{Typical velocity of monopoles in the intergalactic space, due to acceleration in IGMFs. The product of the velocity magnitude and Lorentz factor is shown as a function of monopole mass $m$ and number density $n$.
The monopole charge is fixed to a single Dirac charge, while
the IGMF amplitude is fixed at $\bar{B} = 10^{-15}$~G (\textit{\bfseries left panel}) and $\bar{B} = 10^{-9}$~G (\textit{\bfseries right panel}).
The gray line indicates where $\Omega(m,n) \sim H_0$. Above this line,  the $\gamma v$ amplitude is given by Eq.~\eqref{gamma_v_max}, while below it (in the gray shaded region) it is estimated using Eq.~\eqref{gamma_v_max_hubble}.
}
    \label{fig:oscillation_gammav}
\end{figure*}

The product of the monopole velocity and Lorentz factor, $\gamma v$, is a useful quantity in both the relativistic and non-relativistic regimes.
Its oscillation amplitude during the magnetic plasma oscillations is written using (\ref{energy-conserv}) as
\begin{equation}
    \overline{\gamma v} = \frac{\bar{B}^2}{2mn} \sqrt{1 + \frac{4mn}{\bar{B}^2}}\,. 
    \label{gamma_v_max}
\end{equation}
This also gives the magnitude of $\gamma v$ averaged over the oscillations, up to order-unity numerical factors. 
Hence if $\Omega \gg H_0$, i.e., if the monopoles undergo many oscillation cycles during a Hubble time, then $\overline{\gamma v}$ represents the 
typical monopole velocity.
In the ultra-relativistic regime ($m n \ll \bar{B}^2$), eq.~(\ref{gamma_v_max}) yields
\begin{equation} 
    \overline{\gamma v}
    \sim 20 \left(\frac{m}{1\text{GeV}} \right)^{-1}\!\left(\frac{n}{10^{-30}\text{cm}^{-3}}\right)^{-1}\! \left(\frac{\bar{B}}{10^{-15}\text{G}}\right)^2,   
\end{equation}
whereas in the non-relativistic regime ($m n \gg \bar{B}^2$), we get 
\begin{equation}
    \overline{\gamma v} \!\sim \! 7\!\cdot\! 10^{-5}\!\left(\!\frac{m}{10^{10}\text{GeV}}\!\right)^{\!-\!\frac{1}{2}} \!\! \left(\!\frac{n}{10^{-\!30}\text{cm}^{\!-\!3}}\!\right)^{\!-\!\frac{1}{2}} \!\! \left(\!\frac{\bar{B}}{10^{-\!15}\text{G}}\!\right).
\end{equation} 

If on the other hand $\Omega \ll H_0$, 
the monopoles do not complete a full oscillation period within a Hubble time and hence do not reach the maximum value $\overline{\gamma v}$.
In other words, the backreaction of the monopoles on the IGMF is negligible.
Hence in a homogeneous IGMF,\footnote{Here we are actually assuming the IGMF to have a sufficiently large coherence length such that the monopoles are linearly accelerated over a Hubble time. For acceleration in IGMFs with smaller coherence lengths, see \cite{Perri_2024}.} 
 $\gamma v$ grows linearly in time (cf. (\ref{mm-eom})), and in a Hubble time its magnitude becomes
\begin{equation}
    (\gamma v)_H \sim \frac{g\bar{B}}{m H_0} \sim 3\times 10^{-2}\frac{g}{\bar{g}} \left(\frac{m}{10^{10} \,\text{GeV}}\right)^{-1} \left(\frac{\bar{B}}{10^{-15}\, \text{G}}\right),
    \label{gamma_v_max_hubble}
\end{equation}
in both relativistic and non-relativistic cases.

The typical velocity of monopoles in the intergalactic space,
given by \eqref{gamma_v_max} for $\Omega \gg H_0$ 
and (\ref{gamma_v_max_hubble}) for $\Omega \ll H_0$,
can collectively be written as 
\begin{equation}
    \gamma v \sim \min \left\{\overline{\gamma v}, \, (\gamma v)_H  \right\}\,.
    \label{gamma_v_tot}
\end{equation}
We show its dependence on monopole parameters in \fref{fig:oscillation_gammav}. The velocity is governed by Eq.~\eqref{gamma_v_max} above the $\Omega \sim H_0$ line and by Eq.~\eqref{gamma_v_max_hubble} below it.

\subsection{Monopole displacement and inter-monopole distance}
\label{app:C2}

One characteristic spatial scale of monopole magnetic plasma oscillations, under the assumption of a spatially uniform magnetic field, is the
displacement of the individual monopoles during the oscillations, as derived in (\ref{eq:d}). This can be rewritten as
\begin{equation}
     d = \frac{\bar{B}}{g n} \sim \left(\frac{g}{\bar{g}}\right)^{-1} \left(\frac{n}{10^{-30} \, \text{cm}^{-3}}\right)^{-1} \left(\frac{\bar{B}}{10^{-15}\, \text{G}}\right)\, \text{kpc} \,.
    \label{travel-distance}
\end{equation}

\begin{figure}[t]
    \centering
    \includegraphics[width=0.9\linewidth]{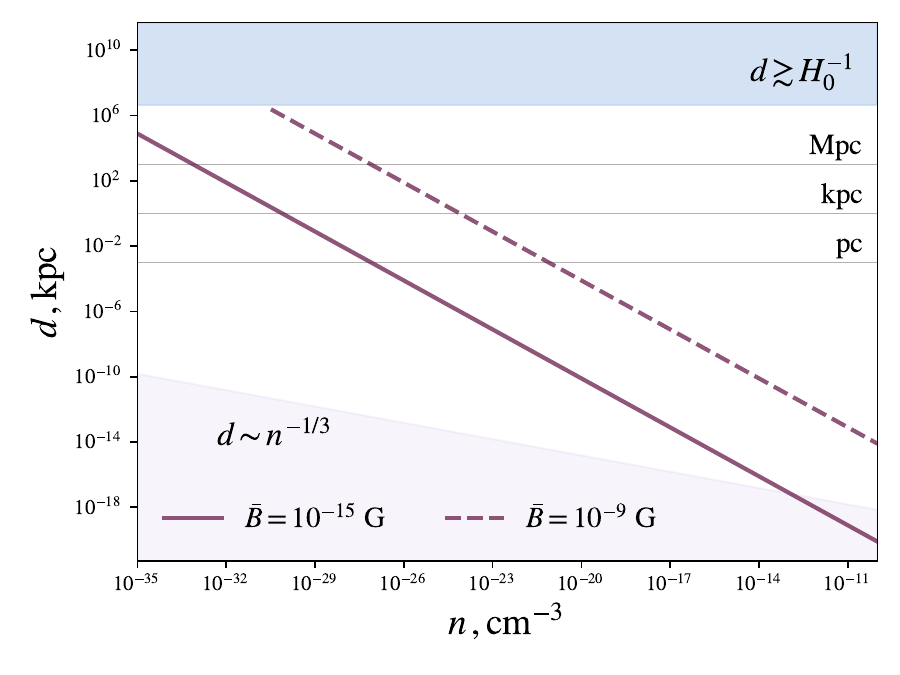}
    \caption{Spatial displacement of individual monopoles during plasma oscillations, as a function of monopole number density.
The monopole charge is fixed to a single Dirac charge, while 
the IGMF strength is set to $\bar{B} = 10^{-15}$~G and $\bar{B} = 10^{-9}$~G. 
The continuum treatment is invalid in the pink shaded region, since at its 
upper edge the displacement is comparable to the inter-monopole distance.
The gray shaded region corresponds to distances exceeding the Hubble radius.
    }
    \label{fig:oscillation_distance}
\end{figure}

In the main part of this paper we modeled the monopole population as a continuum, and used density and bulk velocity fields to analyze the plasma oscillations. For this treatment to be valid, the characteristic scale of the oscillation~$d$ should be larger than the inter-monopole distance, i.e.,
\begin{equation}
 d \gtrsim n^{-1/3}. 
\label{eq:C.10}
\end{equation}
This translates into a condition on the monopole density:
\begin{equation}
 n \lesssim \left( \frac{\bar{B}}{g}  \right)^{3/2}
\sim 10^{-13}\, \ro{cm}^{-3}
\left( \frac{g}{\bar{g}} \right)^{-3/2}
\left( \frac{\bar{B}}{10^{-15}\, \mathrm{G}} \right)^{3/2}.
\label{eq:31}
\end{equation}
This turns out to be an upper limit, because the inter-monopole distance~$n^{-1/3}$ decreases with $n$ more slowly than $d\propto n^{-1}$.
The condition (\ref{eq:31}) is equivalent to requiring that the force a monopole feels from its neighboring monopoles be weaker than that from the IGMF, as is assumed in the monopole equation of motion~(\ref{eom-covariant}). 
In fact, a Coulomb-like magnetic field around a monopole, $B \sim g/r^2$, measured at the average inter-monopole distance, $r \sim n^{-1/3}$, is smaller than the IGMF if $g n^{2/3} \lesssim \bar{B}$.
We plot the displacement scale in \fref{fig:oscillation_distance}, where one clearly sees that (\ref{eq:31}) is violated only at rather large monopole densities.

The upper limit (\ref{eq:31}) on the monopole density is generically larger than our blazar limit~(\ref{eq:n_upper_bound}) (unless going to ultraheavy masses beyond what are shown in Fig.~\ref{fig:project_sensitivity}). This justifies the continuum approach. A much higher monopole density can invalidate the continuum treatment and hence may be unconstrained by blazars; however such densities are ruled out by other experiments, as shown in the right panel of Fig.~\ref{fig:project_sensitivity}.

Note also that since monopoles travel slower than light, it follows that $\Omega^{-1} \gtrsim d$, which combined with (\ref{eq:C.10}) implies
$\Omega^{-1} \gtrsim n^{-1/3}$. If, furthermore, monopoles are sufficiently abundant such that they affect the deflection of the blazar cascade,
i.e. $D_e > \Omega^{-1} $, then the distance traveled by the electrons/positrons is necessarily larger than the mean inter-monopole distance, 
i.e. $D_e \gtrsim n^{-1/3}$.\footnote{Likewise, if the plasma oscillation finishes one cycle within the Hubble time, $\Omega > H_0$, then (\ref{eq:C.10}) guarantees that the inter-monopole distance is smaller than the Hubble radius, $ n^{-1/3} \lesssim H_0^{-1}$.}

%--------------------
%  Fig 12
%--------------------

\section{Spatial inhomogeneities in monopoles-IGMF system}
\label{sec:inhomo}

In this appendix we carry out a very crude estimate of the effect of inhomogeneities in a monopole-magnetic field system
described by (\ref{eq:Max13}) and (\ref{eq:vel16}).
We assume that there is only one time scale and one spatial scale,
and hence rewrite the spacetime derivatives as
$\partial_t \to \Omega$ and $\partial_i \to k$.
We also ignore the vector components, and 
rewrite all fields in terms or their oscillation amplitudes:
$B^i \to \bar{B}$, $E^i \to \bar{E}$, etc.
Further assuming that there is 
one monopole species and no electrically charged particles, 
the evolution equation for the monopoles' velocity field (cf. (\ref{eq:vel16})), together with Faraday and Amp\`ere laws (cf. third and fourth lines of (\ref{eq:Max13})), imply the following parametric relations    
\begin{align}
    m (\Omega + k \bar{v}) \bar{\gamma} \bar{v}
    &\sim g \left(\bar{B} + \bar{v} \bar{E}\right), \\
  k \bar{E} &\sim  \Omega \bar{B} +  g \bar{n} \bar{v}, \\
 k \bar{B} &\sim \Omega \bar{E} .
\end{align}
One can check that if 
$\bar{B}^2 \ll m \bar{n}$ and $ k \ll g \sqrt{\bar{n} / m}$,
then the above set of equations give,
\begin{equation}
 \bar{v} \sim \frac{\bar{B}}{\sqrt{m \bar{n}} } \ll 1 , 
\quad
\Omega \sim g \sqrt{ \frac{\bar{n}}{ m}  } \gg k.
\end{equation}
On the other hand if 
$\bar{B}^2 \gg m \bar{n}$ and $ k \ll g \bar{n} / \bar{B}$, then 
\begin{equation}
 \bar{\gamma} \sim \frac{\bar{B}^2}{m \bar{n}} \gg 1,
\quad
\Omega \sim \frac{g \bar{n}}{\bar{B}}
\gg k.
\label{inhomog-ext}
\end{equation}
We have thus reproduced the nonrelativistic and relativistic solutions for
the velocity amplitude (\ref{energy-conserv}) and the oscillation frequency
(\ref{magnetic-plasma-frequency}),
which were derived for homogeneous fields in \sref{sec:magn-oscillations}.
The results here suggest that the homogeneous analysis is valid if 
the scale of inhomogeneities satisfies $k \ll \Omega$.\footnote{For inhomogeneities with larger $k$, the treatment in this appendix based on a single scale does not necessarily hold.} 
In other words, 
the system can be treated as effectively homogeneous 
as long as the inhomogeneity length scale of the magnetic field is as large as
\begin{align}
    \lambda &\gg \frac{1}{\Omega } \sim \max \left\{\frac{\bar{B}}{g\bar{n}}; \sqrt{\frac{m}{g^2\bar{n}}} \right\} \notag \\ 
     &\sim\, 1\, \text{kpc} \times \max \left\{ \, \left( \frac{g}{\bar{g}}\right)^{-1} \left( \frac{\bar{n}}{10^{-30} \text{cm}^{-3}}\right)^{-1} \left( \frac{\bar{B}}{10^{-15} \text{G}}\right); \right. \notag \\
    &\hspace{4mm}\left. \left( \frac{g}{\bar{g}}\right)^{-1} \left( \frac{m}{100\, \text{GeV}}\right)^{1/2}\left( \frac{\bar{n}}{10^{-30}\, \text{cm}^{-3}}\right)^{-1/2} 
    \right\}\,.
    \label{coherence-length}
\end{align}

Throughout this paper, we have assumed the correlation length of the IGMF to be as large as $\lambda \gg D_e \sim 100\, \ro{kpc}$,
so that the IGMF is effectively homogeneous for the electrons and positrons in the cascade. 
Under this assumption, and noting that monopoles affect the deflection angle of the electrons only if $D_e > \Omega^{-1} $ (cf. (\ref{deflection-angle})), one immediately sees that in the monopole parameter space constrained by blazars, 
the condition~(\ref{coherence-length}) is necessarily satisfied. 
This can also be seen by substituting into
\eqref{coherence-length} the 
monopole density at its blazar-induced upper limit \eqref{eq:n_upper_bound}, which gives
\begin{align}
    \lambda &\gg  \lambda_{\mathrm{UB}} \sim \frac{\tau\theta_\text{psf} E_e}{\bar{e}\bar{B}} \notag \\
     &\sim 26\, \text{kpc} \times \left(\frac{\tau\theta_\text{psf}}{7\times 10^{-3}}\right)\left(\frac{E_{\gamma,s}}{39\,\text{GeV}}\right)^{1/2} \left(\frac{\bar{B}}{10^{-15}\,\text{G}}\right)^{-1}\,.
     \label{coherence-length-upper-bound}
\end{align}

We hence conclude that as long as the IGMF correlation length satisfies 
$\lambda \gg D_e$, our blazar limits on monopoles are valid.
For smaller correlation lengths, the inhomogeneities can modify the bounds on the IGMF strength and/or the monopole abundance.\footnote{Even if the IGMF is initially produced as a homogeneous field, structure formation can induce magnetic field components with finite correlation lengths of the size of cosmic voids, typically of order $10$--$100$~Mpc. This however is much larger than $D_e$.} 

Let us also remark that under the condition (\ref{coherence-length}), 
the phase velocity of the oscillating IGMF,
$v_{\ro{ph}} = \lambda \Omega / 2 \pi$, is larger than unity. 
The phase velocity is thus necessarily larger than the random velocity of the individual monopoles obtained e.g. by scatterings with CMB photons. This guarantees that the plasma oscillation avoids Landau damping, in the parameter space constrained by blazars.

\section{Dissipation}
\label{app:dissipation}

Throughout the main text, we assumed that energy dissipation by magnetic monopoles has a negligible effect on the total energy of the monopole–magnetic field system, which can therefore be treated as conserved over time.

In this appendix, we estimate the rate of magnetic field dissipation induced by magnetic monopoles through several mechanisms, including inverse Compton scattering on the CMB, Larmor radiation in magnetic fields, and energy losses due to Coulomb-like scattering with the intergalactic medium (IGM). We also discuss the potential impact of such dissipative processes on constraints on monopole abundance derived from TeV blazar observations.

\subsection{Inverse Compton scattering of CMB photons}
\label{app:ic-dissipation}

We begin by assessing the energy loss of monopoles via inverse Compton (IC) scattering of CMB photons.
The IC radiation power of magnetic monopoles is analogous to that of electrically charged particles and, in the Thomson regime 
\begin{equation}
\gamma \epsilon_\text{CMB} \ll m, 
\label{eq:T-reg}
\end{equation}
it is given by \cite{Osborne1970}:
\begin{equation}
    P_\text{IC}  = \frac{4}{3} \sigma_{T}^{\scriptscriptstyle{M}} u_{\text{CMB}} \gamma^2 v^2 \,,
    \label{monopole-ic-power}
\end{equation}
where $\sigma_{T}^{\scriptscriptstyle{M}} = \sigma_{T} (g/\bar{e})^4 (m_e/m)^{2}$ is the magnetic Thomson cross section \cite[see, e.g.,][]{Osborne1970, Milton_2006, Baines_2018}.
The rate of energy density loss of the monopole–IGMF system due to IC radiation is
\begin{equation}
    \Gamma_{\ro{IC}} = \frac{2 n P_{\ro{IC}} }{\bar{B}^2}\,.
    \label{IC-rate}
\end{equation}
We further introduce the ratio of the IC dissipation rate to the Hubble expansion rate,
\begin{equation}
    \Delta_\text{IC} = \frac{\Gamma_{\ro{IC}}}{H_0}\,.
    \label{delta-ic}
\end{equation}
If $\Delta_\text{IC} \gtrsim 1$, then it indicates that the energy dissipation significantly affects the system over a Hubble time. 

In this appendix we also consider monopoles with very low densities such that the plasma oscillation does not finish one cycle within the age of the universe. We hence use the general expression (\ref{gamma_v_tot}) for the velocity of monopoles in the intergalactic space. 
If $\Omega > H_0$, and further if the monopole oscillations are relativistic, then the Lorentz factor scales with the number density as
$\gamma \sim \bar{\gamma} \propto n^{-1}$, and thus 
$\Delta_\text{IC} \propto n^{-1}$. In this regime $\Delta_\text{IC} \gtrsim 1$ translates into an upper limit on the number density, whose explicit form is 
\begin{multline}\label{n-ic-boundary-upper}
    n \lesssim  \frac{2}{3} \frac{\eta \bar{B}^2}{m^4} \\ \sim 5\times 10^{-34} \,\text{cm}^{-3}\left(\frac{g}{\bar{g}}\right)^4 \left(\frac{m}{1\,\text{GeV}}\right)^{-4} \left(\frac{\bar{B}}{10^{-15}\,\text{G}}\right)^2\,,
\end{multline}
with $\eta = (g/\bar{e})^4 (\sigma_T m_e^2 u_\text{CMB} / H_0)$. 
If instead the monopole oscillations are nonrelativistic, then $v \sim \bar{v} \propto n^{-1/2}$. In this case $\Delta_\text{IC} $ is independent of~$n$, and $\Delta_\text{IC} \gtrsim 1$ gives an $n$-independent limit on~$m$. 

On the other hand if $\Omega < H_0$, then $\gamma v \sim (\gamma v)_H$, which is independent of $n$. Consequently $\Delta_\text{IC} \propto n$, 
and $\Delta_\text{IC} \gtrsim 1$ translates into a lower limit on~$n$,
\begin{equation}\label{n-ic-boundary-lower}
    n \gtrsim  \frac{3}{8} \frac{m^4 H_0^2}{g^2 \eta} \sim 2 \times 10^{-41}\, \text{cm}^{-3} \left(\frac{g}{\bar{g}}\right)^{-6} \left(\frac{m}{1\,\text{GeV}}\right)^4\,. 
\end{equation}
We stress that the above discussion only applies in the Thomson regime
(\ref{eq:T-reg}).\footnote{We adopt (\ref{eq:T-reg}) as a rough criterion for the applicability of Thomson scattering, however the actual condition may be stricter for monopoles since their size $\sim g^2 / m$ is larger than their Compton wavelength.
We also note that scatterings of solitonic monopoles in the regime where (\ref{eq:T-reg}) is violated may lead to a temporal symmetry restoration, and thus to monopole-antimonopole pair productions.} This condition, upon using \eqref{gamma_v_tot}, can be rewritten as
\begin{multline} \label{m-thomson}
    m \gtrsim \min \left\{\sqrt{\frac{\bar{B}^2\epsilon_\text{CMB}}{2n}} ; \sqrt{\frac{g\bar{B}\, \epsilon_\text{CMB}}{H_0}}\right\}  \sim 
    4\, \text{GeV} \times \\ \min \left\{\left(\frac{n}{10^{-32}\,\text{cm}^{-3}}\right)^{-\frac{1}{2}} \!\left(\frac{\bar{B}}{10^{-10}\,\text{G}}\right);\, 
\left(\frac{g}{\bar{g}}\right)^{\frac{1}{2}}
\left(\frac{\bar{B}}{10^{-10}\,\text{G}}\right)^\frac{1}{2}\right\},
\end{multline}
where the first branch corresponds to the regime of $\Omega > H_0$, and the second to $\Omega < H_0$.
Here we have focused on masses $m \gg \epsilon_{\ro{CMB}} \sim 10^{-3}\, \ro{eV}$; then (\ref{eq:T-reg}) could be saturated only if $\gamma \gg 1$, and so we substituted into $\gamma$ its relativistic expressions. 

\begin{figure*}[t]
    \centering
    \includegraphics[width=\linewidth]{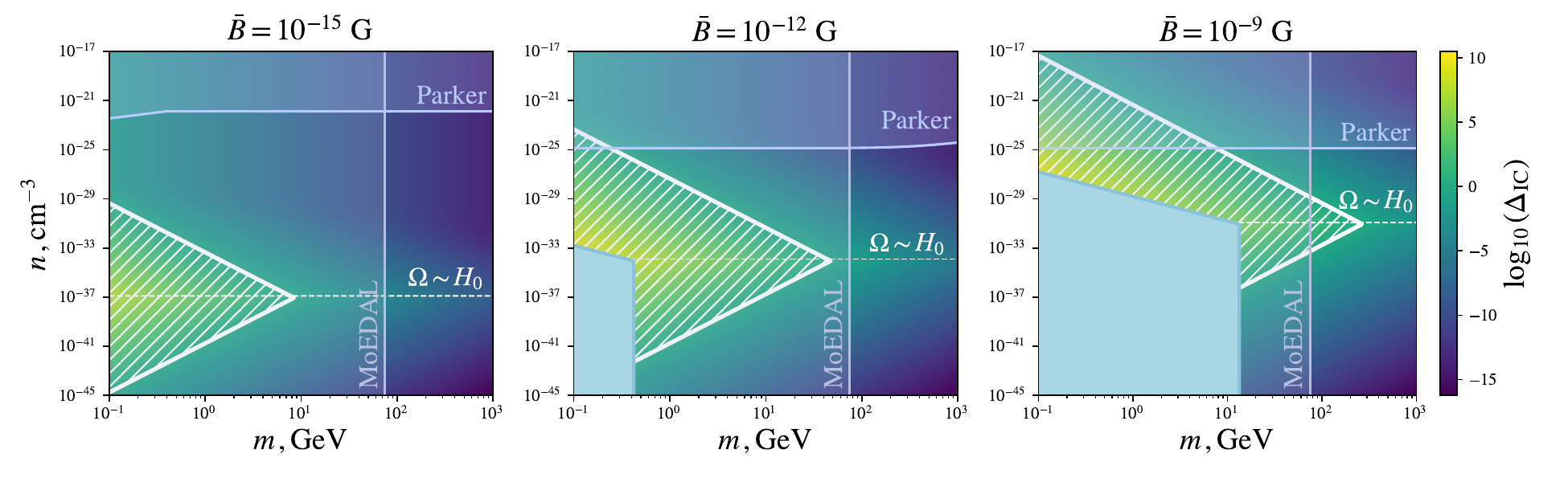}
    \caption{
    The IC dissipation rate $\Delta_\text{IC}$, as defined in \eqref{delta-ic}, is indicated by color. We show the monopole mass $m$ in the range $10^{-1}$ to $10^{3}\,\text{GeV}$ and the monopole number density $n$ in the range $10^{-45}$ to $10^{-17}\,\text{cm}^{-3}$. The monopole magnetic charge is fixed to $g = \bar{g}$, while the IGMF strength is set to $\bar{B} = 10^{-15}$, $10^{-12}$, and $10^{-9}\,\text{G}$, as indicated in the plot titles. The white hatched region indicates where $\Delta_\text{IC} > 1$, i.e.,
IC scatterings of CMB photons significantly dissipates the energy of the monopole-IGMF system during a Hubble time.
The boundaries of this region, shown by white solid lines, are given by \eqref{n-ic-boundary-upper} and \eqref{n-ic-boundary-lower}. The region where the Thomson approximation is not applicable is blocked by a blue non-transparent color, with its boundary given in \eqref{m-thomson}. For comparison, we also show the MoEDAL bound and the galactic Parker bound.
The white dashed line indicates where $\Omega \sim H_0$.
    }
    \label{fig:ic-energy-loses}
\end{figure*}

The monopole parameter space affected by IC dissipation is shown in fig.~\ref{fig:ic-energy-loses}, 
where we have also indicated the region where the Thomson approximation is invalid.
The IC dissipation is efficient for low-mass monopoles $m \lesssim (10 - 300)\,\text{GeV}$, with the detailed upper limit depending on the IGMF strength. However, most of this mass range is ruled out by the MoEDAL constraint, $m \gtrsim 75\,\text{GeV}$.

In the figure we also plotted the galactic Parker bound 
$F \lesssim 3 \times 10^{-16}~\text{cm}^{-2}~\text{s}^{-1}~\text{str}^{-1}$
\cite{Parker:1970xv,Turner_1982}, translated into a bound for the monopole number density using~\eqref{n-to-flux} and \eqref{n-to-flux:velocity}.\footnote{Considering monopoles that do not get trapped inside the Milky Way and
a homogeneous IGMF, the general relation between the monopole flux on Earth and the number density in the intergalactic space is
$F = n v_{\ro{MW}} / 4 \pi$, where
$v_{\ro{MW}} \sim \max (v, v_p)$ with the intergalactic monopole velocity~$v$ given by (\ref{gamma_v_tot}). 
However at densities saturating the galactic Parker limit, $\Omega > H_0$ is realized and hence one can set $v \sim \bar{v}$.}
The relative velocity between monopoles and the Milky Way galaxy $v_{\ro{MW}}$ ranges from
$v_p \sim 10^{-3}$ up to $\bar{v} \sim 1$, which is why the Parker bound on $n$ is seen to vary by three orders of magnitude.

%--------------------
%  Fig 8
%--------------------

\subsection{Larmor radiation}

Magnetic monopoles placed in a magnetic field experience acceleration and therefore emit Larmor radiation, similarly to accelerating electrically charged particles. 
To evaluate the radiation emission power of a monopole, 
we use the relativistic version of the Larmor formula and replace the electric charge with the magnetic:
\begin{equation}
    P_\text{L} = \frac{g^2\gamma^6}{6\pi}\left((\dot{\boldsymbol{v}})^2  - \left(\boldsymbol{v} \times \dot{\boldsymbol{v}}\right)^2\right)\,,
\end{equation}
where an overdot denotes $d/dt$.
For monopoles traveling along a homogeneous IGMF, the acceleration and velocity vectors are parallel, i.e.
$\dot{\boldsymbol{v}} \parallel \boldsymbol{v}$. 
Then using \eqref{mm-eom}, the emission power becomes:
\begin{equation}
    P_\text{L} = \frac{g^2\gamma^6 (\dot{\bd{v}})^2 }{6\pi}= \frac{g^4 \bd{B}^2}{6\pi\, m^2}\,.
\end{equation}
The energy dissipation rate due to Larmor radiation is given by $\Gamma_\text{L} = 2n P_\text{L} /\bar{B}^2$, and its ratio to the Hubble rate is:
\begin{equation}
     \Delta_\text{L} = \frac{\Gamma_L}{H_0} \sim \frac{1}{3\pi} \frac{g^4 n}{m^2 H_0} \,,
    \label{larmor-dissipation-rate}
\end{equation}
where we substituted $\abs{\bd{B}} \sim \bar{B}$ as the typical IGMF strength
upon moving to the far right-hand side. 
Hence we find that the Larmor dissipation has a significant impact,
$\Delta_L \gtrsim 1$, if the monopole density is as large as
\begin{equation}
n \gtrsim 
3\pi \frac{m^2 H_0}{g^4}
\sim
10^{-5} \text{cm}^{-3}
 \left(\frac{g}{\bar{g}}\right)^{-4} \left(\frac{m}{1 \text{GeV}}\right)^{2}.
\end{equation}
Hence in the monopole parameter space not yet excluded by laboratory or astrophysical constraints, Larmor energy losses are negligible.

\subsection{Coulomb-like scatterings of IGM}\label{app:coulomb}

We now estimate the energy dissipation in the monopole–magnetic field system due to Coulomb-like scatterings between monopoles and electrically charged particles in the IGM. These energy losses are dominated by electrons in the plasma, while contributions from protons and ions are suppressed by a factor of $m_e / m_{p,i}$. The corresponding energy loss rate is given by \cite{Perri_2024}:\footnote{In App.~B of \cite{Perri_2024}, the drag force on monopoles due to scatterings with nonrelativistic electrically charged particles was derived in the form
$m d (\gamma \bd{v})/ dt \supset -(P_C / v^2)  \bd{v}$.
Using $d \gamma / d t = \bd{v} \cdot d (\gamma \bd{v}) / dt$, one sees that $P_C$ corresponds to a monopole's energy loss per unit time.}
% \begin{equation}
%     P_C = \frac{e^2 g^2 n_e}{4\pi m_e v}  \left( \ln\left(\frac{2 m_e v}{\omega_\text{pl, e}} \sqrt{\gamma^2 - 1} \right) - \frac{v^2}{2}\right)\,,
% \end{equation}
\begin{equation}
    P_C = C \frac{\bar{e}^2 g^2 n_e v}{4\pi m_e}\,.
\label{eq:P_C}
\end{equation}
Here $C$ is a numerical coefficient roughly of order unity which we omit hereafter, and $v$ is the monopole's velocity magnitude.
This expression is valid given that the monopole velocity is much larger than the thermal velocity of the individual electrons in the rest frame of the plasma, i.e.,
\begin{equation}
v > v_e^\text{th} \sim \sqrt{\frac{T_e}{m_e}},
\label{eq:T_e}
\end{equation}
where $T_e$ is the electron temperature.
The energy loss rate for a monopole is similar to that for a cosmic ray proton (see, e.g., \cite{Leite_2017}) up to the replacement $\bar{e}^4 \to \bar{e}^2 g^2 v^2$ .

The energy loss rate of the monopole-IGMF system 
can be related to that of a single monopole as $\Gamma_C = 2 n P_C / \bar{B}^2$, and its ratio to the Hubble rate is
\begin{equation}
    \Delta_C = \frac{\Gamma_C}{H_0} \sim  \frac{\bar{e}^2 g^2 n_e n v}{2\pi m_e H_0 \bar{B}^2} .
\end{equation}
The condition for the energy dissipation to the IGM to be siginificant, $\Delta_C \gtrsim 1$, translates into a lower limit on~$n$. 
We assume that $n$ at this lower limit is large enough to give $\Omega > H_0$, and verify this assumption \textit{a posteriori}.
Thus using $v \sim \bar{v} \sim \min(1, \bar{B}/\sqrt{mn})$, we obtain
\begin{multline}
    n \gtrsim \max \left\{\frac{1}{A}; \frac{m}{A^2\bar{B}^2} \right\}  \sim 10^{-27}\text{cm}^{-3} \left(\frac{\bar{B}}{10^{-15}\,\text{G}}\right)^2 \times
\\ \max \left\{
\left(\frac{g}{\bar{g}}\right)^{-2}
\!\!\!; 
\left(\frac{g}{\bar{g}}\right)^{-4}
\left(\frac{m}{10^{-1}\,\text{GeV}}\right)\!
\right\},
\label{eq:E15}
\end{multline}
where $A = \bar{e}^2 g^2 n_e / (2\pi m_e H_0 \bar{B}^2)$,
and upon moving to the second line we used 
$n_e \sim 1 \, \ro{m}^{-3}$ for the electron density in the IGM.
We thus see that $n$ satisfying the lower limit indeed gives 
$\Omega > H_0$ (cf. Fig.~\ref{fig:oscillation_freq}). 

On the other hand, the condition (\ref{eq:T_e}) gives an upper limit on $n$, beyond which our estimate becomes invalid. 
Considering again that $\Omega > H_0$ at the upper limit, and moreover since the threshold velocity is $v_e^\text{th} \ll 1$,
we use the nonrelativistic expression
$v \sim \bar{v} \sim \bar{B} / \sqrt{mn}$ to rewrite (\ref{eq:T_e}) as
\begin{multline}
    n \lesssim \frac{m_e \bar{B}^2}{T_e m}  \\  
\sim 10^{-22} \text{cm}^{-3}\left(\frac{m}{10^{-1}\, \text{GeV}}\right)^{-1}\!\left(\frac{\bar{B}}{10^{-15}\,\text{G}}\right)^2\!\left(\frac{T_e}{10^4 \, \text{K}}\right)^{-1}\!.
\end{multline}

\begin{figure*}[t]
    \centering
    \includegraphics[width=\linewidth]{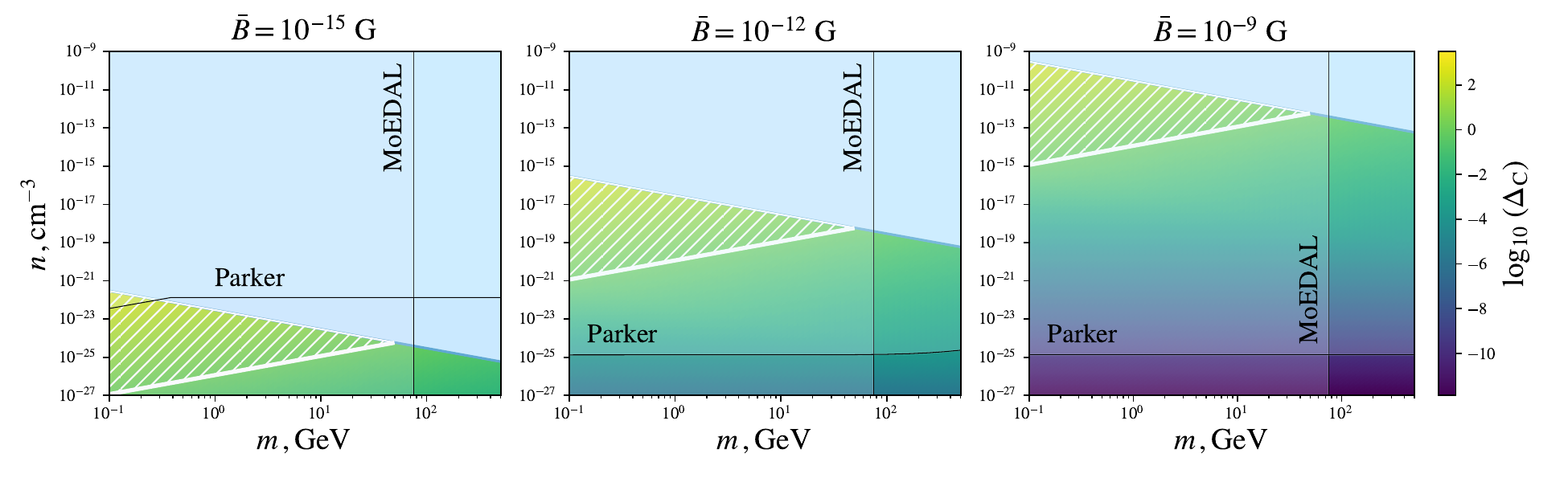}
    \caption{The energy dissipation rate $\Delta_\text{C}$ via Coulomb-like scatterings between monopoles and electrons in the IGM is indicated by color. We show the monopole mass $m$ in the range $10^{-1}$ to $10^{3}\,\text{GeV}$ and the monopole number density $n$ in the range $10^{-27}$ to $10^{-9}\,\text{cm}^{-3}$. We fix the monopole magnetic charge to $g = \bar{g}$, while the IGMF strength is set to $\bar{B} = 10^{-15}$, $10^{-12}$, and $10^{-9}\,\text{G}$, as indicated in the plot titles. For the IGM, we adopt fiducial values of the electron number density $n_e \sim 1\,\text{m}^{-3}$ and temperature $T_e \sim 10^4\,\text{K}$.
    The white hatched region represents the parameter space where Coulomb energy dissipation is efficient, $\Delta_C \ge 1$. The blue blocked region 
    shows where the monopole velocity is smaller than the thermal velocity of electrons, and hence our estimate of the dissipation rate becomes invalid. For comparison, we also show the MoEDAL and galactic Parker bounds.
    }
    \label{fig:coulomb-dissipation}
\end{figure*}

The energy dissipation rate through Coulomb scattering off the IGM is presented in fig.~\ref{fig:coulomb-dissipation}, where we adopt relatively optimistic (i.e., more dissipative) fiducial values for the IGM. Even under these favorable assumptions, Coulomb energy losses become relevant only at rather low masses and high densities, that are mostly excluded by current laboratory and astrophysical constraints.
We also note that the first branch of the $n$~lower limit in (\ref{eq:E15}) appears at mass values smaller than what are shown in the plots.

%--------------------
%  Fig 13
%--------------------

\subsection{Dissipation impact on TeV blazar monopole bound}

The monopole abundance bounds from TeV blazar observations presented in Sec.~\ref{sec:monopole_bound} do not apply if the energy of the monopole–IGMF system is significantly dissipated. 
The most efficient dissipation mechanisms are IC and Coulomb energy losses, as discussed above. Both mechanisms are more efficient at lower monopole masses.

\begin{figure*}
    \centering
    \includegraphics[width=\linewidth]{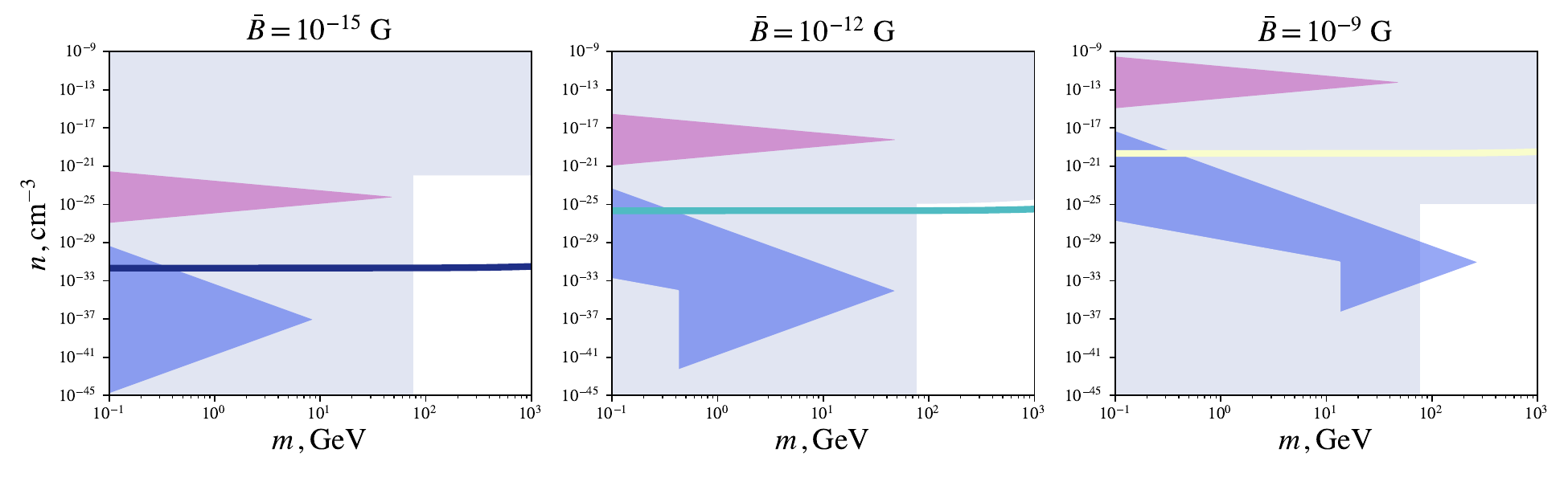}
    \caption{The TeV blazar monopole abundance bounds compared to the parameter space affected by energy dissipation. We show the monopole mass $m$ in the range $10^{-1}$ to $10^{3}\,\text{GeV}$ and the monopole number density $n$ in the range $10^{-45}$ to $10^{-9}\,\text{cm}^{-3}$. 
The monopole magnetic charge is $g = \bar{g}$, while 
the IGMF strength is $\bar{B} = 10^{-15}$, $10^{-12}$, and $10^{-9}\,\text{G}$, as indicated in the plot titles. The TeV blazar bounds from fig.~\ref{fig:project_sensitivity} are shown by solid lines with the same color scheme as in that figure. The parameter space affected by IC dissipation (see Sec.~\ref{app:ic-dissipation} and fig.~\ref{fig:ic-energy-loses}) is shown by a blue shaded region, while the parameter space affected by Coulomb dissipation (see Sec.~\ref{app:coulomb} and fig.~\ref{fig:coulomb-dissipation}) is shown by a purple shaded region. The gray shaded region is excluded by the combined MoEDAL and galactic Parker bounds.
    }
    \label{fig:bounds-vs-dissipation}
\end{figure*}

The TeV blazar monopole bounds are shown together with the parameter space affected by dissipation in fig.~\ref{fig:bounds-vs-dissipation}. 
In the purple region the Coulomb scatterings between monopoles and the IGM are significant and thus the blazar constraint may be evaded, 
however this region is already excluded by the MoEDAL and/or galactic Parker bounds.
In the blue region where IC scatterings of CMB photons are significant, the monopole number densitites are generically lower than those probed by TeV blazars, except for in the upper corner of the region where $m \lesssim 0.3\,\text{GeV}$; this corner may evade the blazar limit, but is excluded by MoEDAL. 

On the other hand with $\bar{B} = 10^{-9}\, \ro{G}$, there is a small window where the IC dissipation is significant and is allowed by the MoEDAL and galactic Parker bounds. 
The dissipation of the IGMF due to monopole energy losses may lead to interesting phenomenology and remains an important topic for future study.

\section{Impact of gravitational potential}
\label{app:external-forces}

The monopole–magnetic-field oscillations are considered in the main text in the absence of non-electromagnetic forces. In this section, we study the effect of the gravitational potential on monopole oscillations. This analysis also generalizes to other external non-electromagnetic forces that satisfy the requirements described below.
As before, we assume a spatially homogeneous scenario, requiring that the external force be uniform, which is a valid assumption far from structures such as massive galaxy clusters. In the presence of an external uniform force $\boldF$, the equations \eqref{maxwell} and \eqref{mm-eom} governing the monopole–magnetic field oscillations modify to
\begin{align}
   m \frac{\partial }{\partial t} \left(\gamma_{\scriptscriptstyle{\text{M}}} \, \boldv_{\scriptscriptstyle{\text{M}}} \right) &= g\boldB + \boldF \,,\\
   m \frac{\partial }{\partial t} \left(\gamma_{\bar{\scriptscriptstyle{\text{M}}}} \, \boldv_{\bar{\scriptscriptstyle{\text{M}}}} \right) &= - g\boldB + \boldF \,,\\
    \frac{\partial}{\partial t} \boldB &=  - \frac{1}{2} gn (\boldvM - \boldvMb)\,,
    \label{external-force-uniform}
\end{align}
where we show the monopoles and anti-monopoles separately. The external force can be decomposed into components parallel and perpendicular to the initial direction of the magnetic field, $\boldF = \boldF_{\parallel} + \boldF_{\perp}$, with $\boldF_{\parallel} \parallel \boldB_i$ and $\boldF_{\perp} \perp \boldB_i$. 
We now study the impact of each component on the magnetic field oscillations separately.

We should note that our estimations of the non-magnetic-field impact remain valid as long as the external force can be approximated as spatially uniform, which should remain valid as long as the distance that the monopoles travel in one oscillation period is much smaller than the gravitational field variation scale, $d \ll \lambda_\text{grav}$, and therefore this approximation is not applicable in close vicinity to massive structures.

\subsection{Parallel gravitational force}

First, we consider the impact of a parallel external force.
Since we neglect the initial monopole velocity, and because monopoles are accelerated by both the magnetic field and the parallel external force, only the parallel component of the monopole velocity is present, $\boldv = \boldv_{\parallel}$. In this scenario, \eref{external-force-uniform} leads to an equation for magnetic field oscillations
\begin{align}
      \frac{\partial^2 \boldB}{\partial t^2}  &=  - \frac{gn}{2} \, \frac{\partial }{\partial t}\left(\boldvM - \boldvMb \right)  \notag \\
      &= - \frac{gn}{2m} \left(\frac{1}{\gammaM^3}\left[g\boldB+ \boldF_{\parallel}\right] - \frac{1}{\gammaMb^3}\left[-g\boldB+ \boldF_{\parallel}\right] \right)  \notag \\ 
      &= -\frac{g^2n}{m\gamma^3}  \boldB\,,
      \label{parrallel-force-B2}
\end{align}
where $1/\gamma^3 \equiv 1/(2\gamma_M^3) + 1/(2\gamma_{\bar M}^3)$. This is valid as long as $\frac{|\boldF_{\parallel}|\Delta\gamma}{g|\boldB|\gamma} \ll 1$, where $\Delta\gamma \equiv \gamma_M - \gamma_{\bar M}$. In the relativistic regime, $\gamma \gg 1$, this condition is satisfied if $|\boldF_{\parallel}| \ll g|\boldB|$. In the non-relativistic regime, $\Delta \gamma \sim v_B v_F$, where $v_B\sim g|\boldB|/(m\Omega)$ is the velocity acquired by monopoles due to acceleration by the magnetic field over one oscillation period, and $v_F\sim |\boldF_{\parallel}|/(mH)$ is the velocity acquired due to acceleration by the external force over a Hubble time.
Therefore, $\frac{|\boldF_{\parallel}|\Delta\gamma}{g|\boldB|\gamma} \sim \frac{|\boldF_{\parallel}|^2}{m^2 \Omega H} \sim v_F^2 H/\Omega$. Since $\Omega \gg H$ in the parameter region of interest, in the non-relativistic regime the equation for magnetic-field oscillations is not modified if $v_F^2 \ll 1$, meaning that the external force is not strong enough to accelerate monopoles to relativistic velocities.  Combining the relativistic and non-relativistic regimes, we obtain the condition on the external-force strength
\begin{equation}
    F \ll \max \left(gB, m\sqrt{\Omega H}\right)\,.
    \label{ext-force-condition}
\end{equation}
Since the primary external force acting on monopoles in extragalactic space is gravity, this condition is satisfied as long as monopoles remain outside of galaxies and are not in the proximity of compact objects. For comparison, the Milky Way galaxy, located on the outskirts of the Virgo Supercluster, has a velocity relative to the CMB of $v_\text{G} \sim 620\, \text{km}/\text{s}$ (see e.g.~\citep{ParticleDataGroup:2024cfk}).

\begin{figure*}[t]
     \centering
     \begin{tabular}{cc}
         \includegraphics[width=0.5\textwidth]{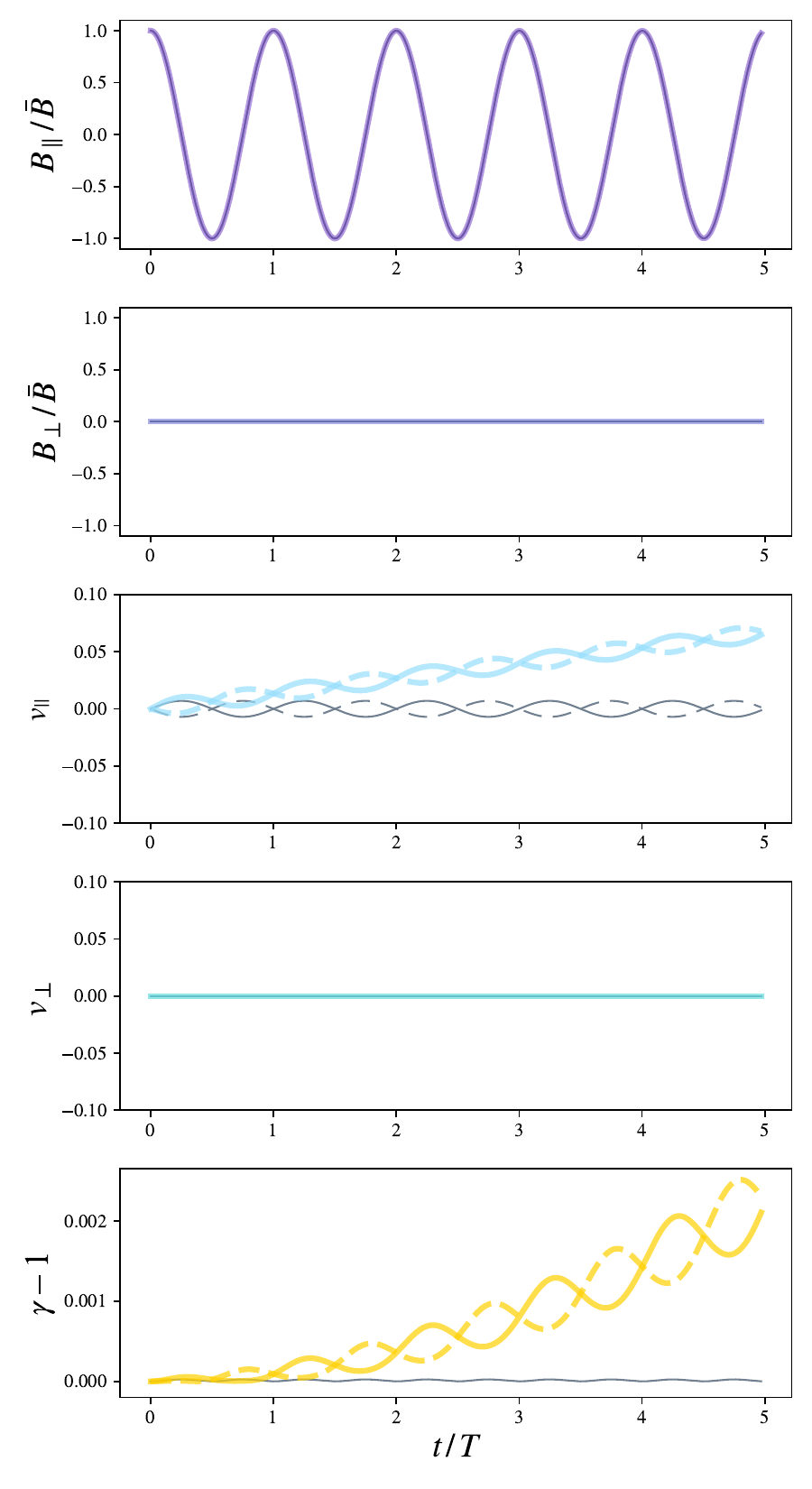} &
         \includegraphics[width=0.5\textwidth]{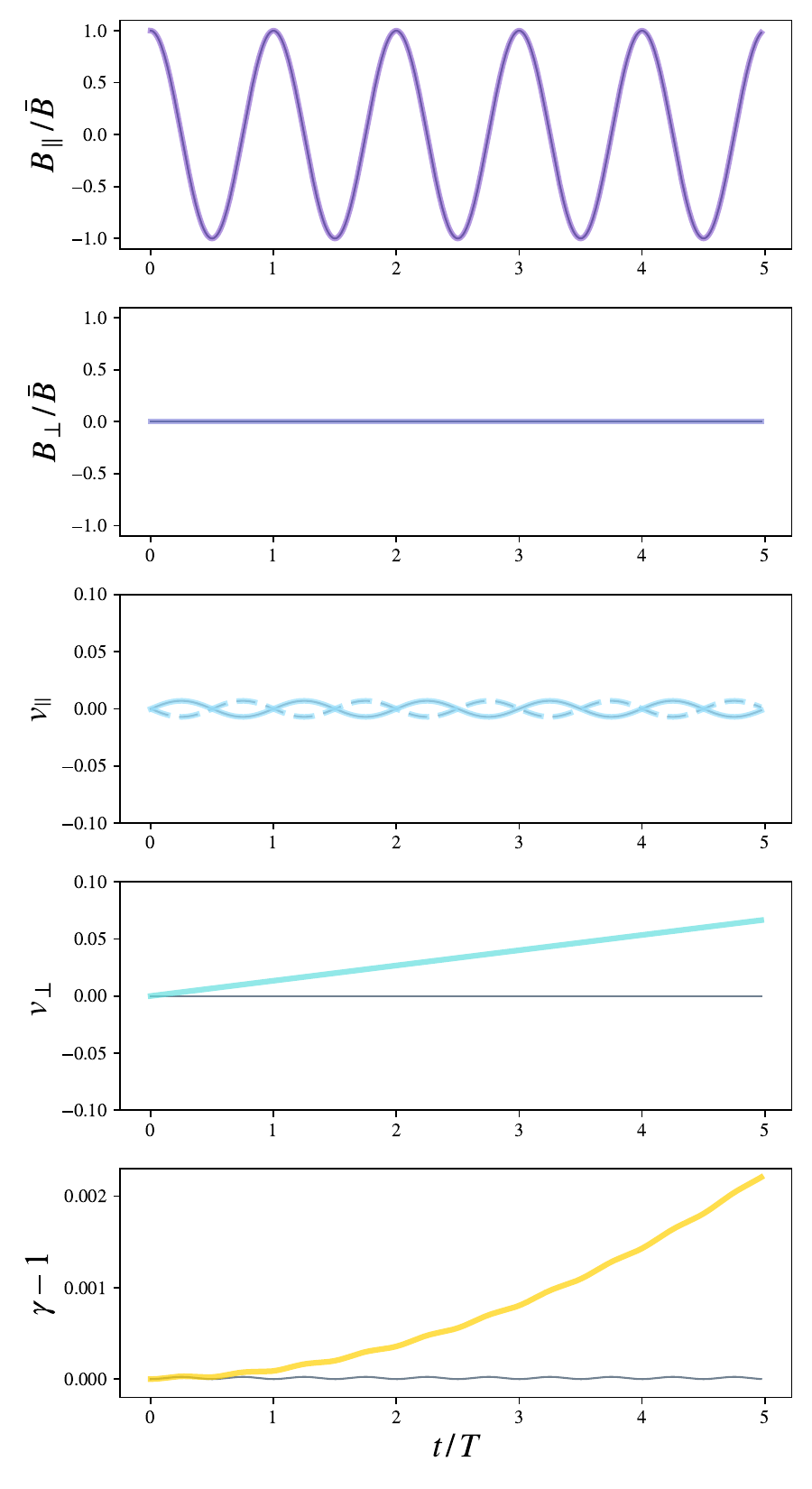}
     \end{tabular}
     \caption{Illustration of monopole–magnetic field oscillations in the presence of a uniform non-electromagnetic 
     force acting on monopoles. In the \textit{\bfseries left panel}, the non-electromagnetic force is directed parallel to the initial direction of the magnetic field, while in the \textit{\bfseries right panel}, the force is perpendicular to the initial direction of the magnetic field. The plots show the numerical solutions of equations \eqref{external-force-uniform} for magnetic monopoles with charge $g = \bar{g}$, mass $m = 10^8$ GeV, and particle number density $n = 10^{-24}\, \text{cm}^{-3}$ in a magnetic field with initial amplitude $\bar{B} = 10^{-11}$ G. 
The uniform non-electromagnetic force is set to be comparable to the gravitational force from a $10^{15} M_\odot$ mass at a distance of $5$ kpc, 
which is significantly stronger than what is expected for monopoles in cosmic voids and is used here for illustrative purposes only. Colored lines show the magnetic-field components parallel and perpendicular to the initial direction, the velocity, and the Lorentz factor of monopoles and antimonopoles over five oscillation periods. Solid lines correspond to monopoles, while dashed lines denote antimonopoles whenever their values differ from those of monopoles. The thin black line behind each main line shows the corresponding result without the non-electromagnetic force.
     }
     \label{fig:oscillations-with-gravity}
\end{figure*}

As we see from \eqref{parrallel-force-B2}, the presence of an uniform parallel external non-electromagnetic force does not affect the magnetic field oscillations, as long as the force is not strong enough to accelerate monopoles to relativistic velocities. The only impact of the external force is the drifting of the monopole and anti-monopole system in the direction of the force, as shown in the left panel of \fref{fig:oscillations-with-gravity}. 

%--------------------
%  Fig 14
%--------------------

\subsection{Perpendicular gravitational force}

Now we consider the impact of an external force perpendicular to the initial direction of magnetic field $\boldF = \boldF_{\perp}$. In this scenario, the monopoles develop both parallel and perpendicular velocity components
$\boldv_{\scriptscriptstyle{\text{M}},\bar{\scriptscriptstyle{\text{M}}}} = \boldv_{\scriptscriptstyle{\text{M}},\bar{\scriptscriptstyle{\text{M}}}} ^{\parallel} + \boldv_{\scriptscriptstyle{\text{M}},\bar{\scriptscriptstyle{\text{M}}}} ^{\perp}$,
and if this leads to the presence of a perpendicular current, a perpendicular component of the magnetic field $\boldB_{\perp}$can also be generated. The time evolution of the parallel and perpendicular components of the magnetic field is governed by the equations:
\begin{align}
    \frac{\partial \boldB_{\parallel}}{\partial t}  &=  -\frac{1}{2} gn (\boldvM^{\parallel} - \boldvMb^{\parallel})\,, \\
    \frac{\partial \boldB_{\perp}}{\partial t}  &=  -\frac{1}{2} gn (\boldvM^{\perp} - \boldvMb^{\perp})\,.
\end{align}
The equations of motion for monopoles become
\begin{align}
      m \frac{\partial}{\partial t} \left(\gamma_{\scriptscriptstyle{\text{M}}} \, \boldv_{\scriptscriptstyle{\text{M}}}^{\parallel} \right) & = g\boldB_{\parallel} \,, \\
      m \frac{\partial}{\partial t} \left(\gamma_{\bar{\scriptscriptstyle{\text{M}}}} \, \boldv_{\bar{\scriptscriptstyle{\text{M}}}}^{\parallel} \right) &=  - g\boldB_{\parallel} \,, \\
      m \frac{\partial}{\partial t} \left(\gamma_{\scriptscriptstyle{\text{M}}} \, \boldv_{\scriptscriptstyle{\text{M}}}^{\perp} \right) &= g\boldB_{\perp} + \boldF_{\perp}\,, \\
      m \frac{\partial}{\partial t} \left(\gamma_{\bar{\scriptscriptstyle{\text{M}}}} \, \boldv_{\bar{\scriptscriptstyle{\text{M}}}}^{\perp} \right) &= - g\boldB_{\perp} + \boldF_{\perp}\,,
\end{align}
where, for both monopoles and antimonopoles, $\frac{\partial}{\partial t} \left(\gamma\, \boldv^{\parallel} \right) = \gamma^3\dot{\boldv}^{\parallel}$ and 
$\frac{\partial}{\partial t} \left(\gamma \, \boldv^{\perp} \right) = \gamma\dot{\boldv}^{\perp} + \gamma^3\boldv^{\perp} \left( \boldv^{\parallel}\cdot\dot{\boldv}^{\parallel}\right)$, assuming $\gammaM \sim \gammaMb \equiv \gamma \sim (1 - v_{\parallel}^2)^{-1/2}$.
This allows us to derive the equations for magnetic field oscillations:
\begin{align}
    \frac{\partial^2 \boldB_{\parallel}}{\partial t^2}  &= - \frac{g^2 n}{m\gamma^3} \boldB_{\parallel} \,,\label{external-force-perpendicular-1} \\
    \frac{\partial^2 \boldB_{\perp}}{\partial t^2}  &= - \frac{g^2n}{m\gamma} \left(\boldB_{\perp} - (\boldvM^{\parallel}\boldB_{\parallel})(\boldvM^{\perp} - \boldvMb^{\perp}) \right)\,. %\notag \\ 
    %& \sim -\frac{g^2n}{m\gamma^3} \boldB_{\perp} \,.
    \label{external-force-perpendicular-2}
\end{align}
To derive the final results, we used $\boldvM^{\parallel} = -\boldvMb^{\parallel}$. Similarly to the previous section, the system of equations \eqref{external-force-perpendicular-1} and \eqref{external-force-perpendicular-2} remains valid if the external force satisfies the condition~\eqref{ext-force-condition}. Since, by definition, the initially perpendicular component $\boldB^\perp_i = 0$, and since we assume $\boldv_{{\scriptscriptstyle{\text{M}}}, i}^{\perp} = \boldv_{\bar{\scriptscriptstyle{\text{M}}},i}^{\perp}$, the perpendicular component of the magnetic field always remains zero.
The numerical solutions for the monopole–magnetic field oscillations in this scenario are shown in the right panel of \fref{fig:oscillations-with-gravity}.
Similarly to the parallel external force, the perpendicular force component does not affect the frequency of the magnetic field oscillations. Instead, it causes the system of monopoles and anti-monopoles to drift uniformly in the direction of the external force.

\section{Impact of arbitrary monopole velocity distribution}
\label{app:velocity-dsitribution}

In the main text we considered the monopoles to move with a uniform velocity. In this appendix we consider a system of monopoles with an arbitrary velocity distribution and analyze whether a non-trivial velocity distribution can affect the monopole-driven magnetic field oscillations. 

We describe the system using the distribution functions 
$f(t, \boldr, \boldp) = dN/d^3r/d^3p$ for the monopoles ($\fM$) and anti-monopoles ($\fMb$), and consider each of them to obey
the collisionless Boltzmann equation:
\begin{equation}
    \frac{\partial f}{\partial t} + \frac{\mathrm{d} \boldr}{\mathrm{d} t} \cdot \frac{\partial f}{\partial \boldr} + \frac{\mathrm{d} \boldp}{\mathrm{d} t} \cdot \frac{\partial f}{\partial \boldp} = 0 \,.
    \label{boltzmann}
\end{equation}
Here $\bd{v} = d \bd{r} / dt$ and $\bd{p} = m \gamma \bd{v}$, with which the Lorentz factor is written as $\gamma = \sqrt{1 + \abs{\bd{p}}^2 / m^2}$. 

As in the main text, we assume spatial homogeneity. Supposing also the absence of electric fields, and using 
$d \bd{p} / d t = \pm g \bd{B}$ with the upper sign for monopoles (charge $g$) and lower for anti-monopoles ($-g$), the Boltzmann equation gives
\begin{align}
    \frac{\partial \fM}{\partial t}  + g \boldB \cdot \frac{\partial \fM}{\partial \boldp} &= 0\,, \\
    \frac{\partial \fMb}{\partial t}  - g \boldB \cdot \frac{\partial \fMb}{\partial \boldp} &= 0\,.
    \label{boltzmann-rel-monopoles}
\end{align}
Moreover from Gauss's law for magnetism and Faraday’s law,
i.e. the second and third lines in (\ref{maxwell-equations}),
we obtain:
\begin{align}
    0 &= \rho_m = g \int \mathrm{d}^3p\, \left(\fM - \fMb \right) \,, \\
    \frac{\partial \boldB}{\partial t} &= - \bd{j}_m  = -g \int \frac{\mathrm{d}^3p}{m \gamma }\, \boldp \left(\fM - \fMb \right) \,.
\label{eq:G5}
\end{align}
The first equation indicates that the net magnetic charge is zero.
Taking a time derivative of both sides of the second equation
and substituting the Boltzmann equations, one finds
\begin{align}
\frac{\partial^2 \boldB}{\partial t^2} &= g^2 \int \frac{\mathrm{d}^3 p}{m \gamma} 
\boldp
\left[
\boldB \cdot \frac{\partial }{\partial \boldp} 
(\fM + \fMb)
\right] \notag \\ 
    &= - g^2 \int \frac{\mathrm{d}^3p}{m \gamma^3}\left\{ \boldB - \frac{\boldp\times[\boldp\times \boldB]}{m^2} \right\}
\left(\fM + \fMb \right) \,,
\label{eq:G6}
\end{align}
where upon moving to the second line we integrated by parts and dropped surface terms. 
The first term in the curly brackets reproduces the standard magnetic field oscillations discussed in the main text, whereas the second term encodes the effect of momentum components transverse to 
$\boldB$. 

In the nonrelativistic limit ($\gamma \to 1$),
one can ignore the second term in the curly brackets in (\ref{eq:G6}) and obtain
\begin{equation}
    \frac{\partial^2 \boldB}{\partial t^2} = - \frac{ g^2 n}{m} \boldB \,,
\end{equation}
where $n$ is the total number density of monopoles and anti-monopoles given by 
\begin{equation}
 n = \int \mathrm{d}^3p\, \left(\fM + \fMb \right).
\end{equation}
Thus, independently of the velocity distribution of the monopoles,
the magnetic oscillation equation takes the same form as for the uniform velocity case studied in the main text. 

For the relativistic case, 
let us focus on a system where the transverse momenta are negligible, and consider vector quantities to have components only along the $z$~direction: 
$\boldB(t) = (0, 0, B_z(t))$, 
$f_{\scriptscriptstyle{\text{M}}, \bar{\scriptscriptstyle{\text{M}}}} 
(t, \bd{p}) = \delta (p_x) \delta (p_y) 
\tilde{f}_{\scriptscriptstyle{\text{M}}, \bar{\scriptscriptstyle{\text{M}}}}  
(t, p_z)$.
Then (\ref{eq:G6}) yields
\begin{equation}
    \frac{\partial^2 B_z}{\partial t^2} =- \frac{g^2 n}{m\langle \gamma^3 \rangle}  B_z \,, 
\end{equation}
where
\begin{equation}
    \frac{1}{\langle \gamma^3 \rangle} \equiv 
\frac{
\int \mathrm{d} p_z\, \gamma^{-3}
\left( 
\tilde{f}_{\scriptscriptstyle{\text{M}}} 
+ \tilde{f}_{ \bar{\scriptscriptstyle{\text{M}}}} 
 \right)
}{
\int \mathrm{d} p_z
\left( 
\tilde{f}_{\scriptscriptstyle{\text{M}}} 
+ \tilde{f}_{ \bar{\scriptscriptstyle{\text{M}}}} 
 \right)
}
\end{equation}
is the expectation value of $\gamma^{-3}$.
This shows that, when transverse momenta are small, the equation describing magnetic field oscillations remains similar to that for monopoles with a uniform velocity distribution, with the only difference being that the Lorentz factor term must now be averaged over the monopole/anti-monopole momentum distribution.

For both the nonrelativistic and relativistic cases, the total energy density of the monopole-magnetic field system is locally conserved,
\begin{equation}
    \mathcal{E} = \frac{\abs{\boldB}^2}{2} + \int \mathrm{d}^3p\, 
m (\gamma - 1)
\left(\fM + \fMb \right) = \text{const} \,.
\end{equation}
For a general momentum distribution the kinetic energies of the individual monopoles and anti-monopoles do not necessarily vanish simultaneously
during the oscillations; in such cases the magnetic field's oscillation amplitude would be smaller than $\sqrt{2 \mathcal{E}}$. 

\bibliographystyle{JHEP}
\bibliography{lib} 

\end{document}